\documentclass[fleqn,usenatbib]{mnras}

\usepackage{newtxtext,newtxmath}

\usepackage[T1]{fontenc}

\DeclareRobustCommand{\VAN}[3]{#2}
\let\VANthebibliography\thebibliography
\def\thebibliography{\DeclareRobustCommand{\VAN}[3]{##3}\VANthebibliography}

\usepackage{graphicx}	
\usepackage{amsmath}	
\usepackage{xcolor}

\newcommand{\WD}{SDSS\,J1143$+$6615}
\newcommand{\WDlong}{SDSS\,J114333.48$+$661531.83}

\newcommand{\Teff}{\mbox{$T_\mathrm{eff}$}}

\newcommand{\jangstrom}{\mbox{erg\,s$^{-1}$\,cm$^{-2}$\,\AA$^{-1}$}}
\newcommand{\Ion}[2]{#1\,\textsc{#2}}
\newcommand{\kms}{\mbox{km\,s$^{-1}$}}
\defcitealias{QCUMBRE}{QCUMBRE}
\defcitealias{cfour}{CFOUR}

\title[A DZ white dwarf with a 30\,MG magnetic field]{A DZ white dwarf with a 30\,MG magnetic field}

\author[M. A. Hollands et al.]{
M. A. Hollands,$^{1}$\thanks{E-mail: m.hollands@sheffield.ac.uk}
S. Stopkowicz,$^{2,3,4}$
M.-P. Kitsaras,$^{3}$
F. Hampe,$^{3}$
S. Blaschke,$^{3}$
and J.J. Hermes$^{5}$
\\
$^{1}$ Department of Physics and Astronomy, University of Sheffield, Sheffield, S3 7RH, UK \\
$^{2}$ Fachrichtung Chemie, Universit{\"a}t des Saarlandes, D-66123 Saarbr{\"u}cken, Germany \\
$^{3}$ Department Chemie, Johannes Gutenberg-Universit\"at Mainz,
Duesbergweg 10-14, D-55128 Mainz, Germany \\
$^{4}$ 
Hylleraas Centre for Quantum Molecular Sciences, Department of Chemistry, University of Oslo, P.O. Box 1033 Blindern, N-0315 Oslo, Norway \\
$^{5}$ Department of Astronomy \& Institute for Astrophysical Research, Boston University,
725 Commonwealth Ave., Boston, MA 02215, USA \\
}

\date{Accepted 2023 January 11. Received 2023 January 11; in original form 2022 November 28}

\pubyear{2023}

\begin{document}
\label{firstpage}
\pagerange{\pageref{firstpage}--\pageref{lastpage}}
\maketitle

\begin{abstract}
Magnetic white dwarfs with field strengths below 10\,MG are easy to recognise
since the Zeeman splitting of spectral lines appears proportional to the magnetic field strength.
For fields $\gtrsim 100$\,MG, however, transition wavelengths become chaotic,
requiring quantum-chemical predictions of wavelengths and oscillator strengths
with a non-perturbative treatment of the magnetic field. While highly accurate calculations have
previously been performed for hydrogen and helium, the variational techniques employed become
computationally intractable for systems with more than three to four electrons.
Modern computational techniques, such as finite-field coupled-cluster theory,
allow the calculation of many-electron systems in arbitrarily strong magnetic fields.
Because around 25\,percent of white dwarfs have metal lines in their spectra,
and some of those are also magnetic, the possibility arises for some metals to be observed in very
strong magnetic fields, resulting in unrecognisable spectra.
We have identified \WDlong\ as a magnetic DZ white dwarf,
with a spectrum exhibiting many unusually shaped lines at unknown
wavelengths. Using atomic data calculated from computational finite-field coupled-cluster methods,
we have identified some of these lines arising from Na, Mg, and Ca. Surprisingly,
we find a relatively low field strength of 30\,MG, where the large number of
overlapping lines from different elements make the spectrum challenging
to interpret at a much lower field strength than for DAs and DBs.
Finally we model the field structure of \WD\ finding the data
are consistent with an offset dipole.
\end{abstract}




\section{Introduction}

The first magnetic white dwarf was discovered by \citet{kempetal70-1}, through
the detection of circularly polarised light from GJ\,742. Since then, many
hundreds of magnetic white dwarfs have been discovered \citep{kawkaetal07-1, kepleretal13-1},
with observed fields strengths spanning a few 10\,kG up to about 1000\,MG.
For fields ranging between a few 100\,kG to a few 10\,MG, magnetic DA white dwarfs (i.e. those with
spectra dominated by hydrogen absorption lines) are easy to identify in intensity spectra
and their field strengths are simple to measure, as many hydrogen lines split into three components,
where the degree of splitting is proportional to field strength. For smaller fields, where such
splitting is unresolved, spectropolarimetry can be used instead
\citep{bagnulo+landstreet18-1,bagnulo+landstreet19-1,bagnulo+landstreet21-1,landstreet+bagnulo19-1}. However, due to reduced throughput, spectropolarimetry is limited to only the brightest white dwarfs.

For higher fields, particularly those beyond 100\,MG, identification is often still straightforward,
though measuring the field strength is no-longer trivial. The diamagnetic
term in the Hamiltonian of the hydrogen atom \citep{wickramasinghe+ferrario00-1}
(resulting in the quadratic Zeeman effect due to its
$B^2$ dependence), quickly exceeds the interaction strength of the paramagnetic term
(linear Zeeman effect), and eventually even the electrostatic potential.
This results in large shifts in wavelength, which
ostensibly appear chaotic in their field strength dependence.
Due to the $n^4$ dependence on the quadratic Zeeman effect
\citep[where $n$ is the principle quantum number,][]{wickramasinghe+ferrario00-1},
the shifts are first observed in the higher order Balmer lines,
but beyond a few 10\,MG also
causes the wavelengths of the H$\alpha$ components to become chaotic.
Because the size of the diamagnetic term in the Hamiltonian becomes comparable to the other terms, and overall the magnetic field is no longer a small perturbation to the system, 
the energies (and hence transition wavelengths), cannot be determined using
perturbation theory, and instead must be determined numerically.

For hydrogen, the first detailed atomistic calculations were performed in the 1980s \citep{roesneretal84-1,forsteretal84-1,henry+oconnell85-1,wunner87-1}.
The results of these calculations quickly found application to assignment of
lines in strongly magnetic white dwarf spectra
\citep{greensteinetal85-1,angeletal85-1,schmidtetal86-2}.
More recent calculations have refined the atomic data for hydrogen in strong
fields \citep{schimeczek+wunner14-1,schimeczek+wunner14-2}.

Even at these early stages, however, the magnetic white dwarf GD\,229
was found to defy assignment of hydrogen spectral lines, leading to speculation
that it may instead have a helium dominated atmosphere
\citep{green+liebert81-1,schmidtetal90-1,schmidtetal96-2}.
This hypothesis was proved correct when the first calculations of \Ion{He}{i}
by \citet{jordanetal98-2} were matched to lines in the spectrum of GD\,229,
implying a surface field varying between $300$--$700$\,MG.
The calculations themselves relied on  
 finite-field full configuration interaction (ff-FCI) theory, a variational technique
providing near-exact solutions to the time-independent electronic Schr\"odinger equation.
Such a description is needed due to electron-electron repulsion term in the Hamiltonian. 
Similar calculations for \Ion{He}{i} were also been performed by
\citet{beckenetal99-1}.

Calculations using variational approaches have been performed for systems with more electrons
such as \Ion{Li}{i} \citep{zhao18-1}, however for systems with more than
three to four electrons, ff-FCI becomes numerically intractable due to
the factorial scaling in computation time.

Fortunately, while white dwarfs with heavy elements in their atmospheres
have been known for more than a century, those with magnetic fields have hitherto
not been observed with field strengths exceeding $\sim 10$\,MG, where atoms are
safely in the Paschen-Back regime. White dwarfs with heavier elements fall into
two main classes: the DQs containing spectral features from carbon, and the DZs containing
features from heavier metals \citep{sionetal83-1} such as calcium and iron.

DQ white dwarfs, those with spectral features from carbon in their atmospheres
(detected from C$_2$ Swan bands at low \Teff\ and \Ion{C}{i/ii} at higher \Teff)
are generally understood to originate from convective dredge up of carbon from
the core into the surrounding helium envelope \citep{fontaineetal84-1,pelletieretal86-1,macdonaldetal98-1},
though a separate population of massive DQs are thought to originate as the product of mergers
\citep{dufouretal07-1,dunlap+clemens15-1,williamsetal16-1,kawkaetal20-1,hollandsetal20-1}.
Of these hot suspected merged DQs, a moderate fraction are also magnetic,
showing Zeeman split \Ion{C}{i/ii} lines -- some with field strengths of a few
MG \citep[e.g.][]{dufouretal08-1}. At lower \Teff\ some peculiar DQs (such as
LHS 2229) show highly distorted and shifted Swan bands which have previously been
hypothesised to arise from strong (100s of MG) magnetic fields. However, \citet{kowalski10-1}
demonstrated that the distorted molecular bands primarily result from pressure-effects
occurring in high-density, low \Teff, helium-dominated white dwarf atmospheres.
To date, no predictions for the wavelengths of atomic or molecular carbon transitions in
strong magnetic fields have been performed.

White dwarfs with metals in their atmospheres are denoted with a Z in their
spectral type, e.g. DAZ, DBZ, or DZ, depending which other lines are visible
in their spectra. DZs specifically (the subject of this work) usually have helium dominated atmospheres,
though are too cool to exhibit \Ion{He}{i} lines ($\Teff < 11,000$\,K), although for
$\Teff < 5000$\,K hydrogen lines are also diminished in strength, and
so in some cases hydrogen atmosphere white dwarfs can also be classed DZ.
Unlike the carbon in DQs, the metals observed in DZs (and DAZs/DBZs etc.)
require an external source, as gravitational settling should deplete white
dwarf atmospheres of metals on timescales that are always much shorter than white
dwarf ages \citep{paquetteetal86-2} --
specifically in the case of cool DZs, sinking timescales are on the order
of $10^{6\text{--}7}$\,yr, whereas their ages range from $10^{9\text{--}10}$\,yr
\citep[see][Figure 1]{wyattetal14-1}.

A vast array of evidence now supports accretion of exoplanetesimals from an
accompanying planetary system as the source of this metal pollution.
Many metal-rich white dwarfs are observed with infra-red excesses resulting from
circumstellar debris disks \citep{zuckerman+becklin87-1,jura03-1,rochettoetal15-1,swanetal19-2},
with a sub-population of those also exhibiting
gaseous emission from the sublimated part of the disk
\citep{gaensickeetal06-3, gaensickeetal07-1, dennihyetal16-1, manseretal20-1, manseretal21-1}.
In a few cases,
when the disk is viewed edge-on, irregular transits are observed demonstrating
the tidal disruption of exoplanetesimals close to the white dwarf Roche radius
\citep{vanderburgetal15-1,vanderboschetal20-1,vanderboschetal21-1, guidryetal21-1, farihietal22-1}.
In two cases the presence of planets themselves has been directly inferred, firstly
from the accretion of an evaporating gas giant by  WD\,J0914+1914 \citep{gaensickeetal19-1}, and
secondly from planetary transits at WD\,1856+534 \citep{vanderburgetal20-1}.
Despite these various sources of evidence for white dwarf planetary systems, white dwarf spectra containing metal lines
remains the most common observable, and can be used to infer the composition
of the accreted exoplanetesimals \citep{zuckermanetal07-1, kleinetal10-1, gaensickeetal12-1, dufouretal12-1, farihietal13-1, xuetal14-1, wilsonetal15-1, hollandsetal17-1, hollandsetal18-2, blouinetal19-1, doyleetal19-1, swanetal19-1, hoskinetal20-1, izquierdoetal21-1, hollandsetal22-1}.
A sub-population of DZs have also been found to exhibit magnetism.

The first discovered magnetic DZ (spectral type DZH) was LHS\,2534 \citep{reidetal01-1},
which was found to have a 1.9\,MG field strength from Zeeman split lines
of \Ion{Na}{i}, \Ion{Mg}{i}, and blended Zeeman components from \Ion{Ca}{i/ii}.
The field strength of LHS\,2534 was recently revised to 2.1\,MG
by \citet{hollandsetal21-1} along with the detection of Zeeman splitting
of \Ion{Li}{i} and \Ion{K}{i}. Since this initial discovery, additional
DZHs were identified by \citet{schmidtetal03-1} and \citet{dufouretal06-1}
(WD\,0155$+$003 and G\,165$-$7, respectively). With the advent of data release 10 (DR10) of the
Sloan Digital Sky Survey (SDSS), \citet{hollandsetal15-1} identified a further seven
objects, bringing the known sample to ten, and finding a high
magnetic incidence of $13\pm4$\,percent for DZs.
With SDSS DR12, \citet{hollandsetal17-1} measured the fields of an additional 15 DZs\footnotemark,
with the range of surface averaged field strengths, $B_s$,
spanning $0.57\pm0.04$\,MG to $10.70\pm0.07$\,MG. Like LHS\,2534, most of these
DZs were identified from Zeeman triplets arising from the \Ion{Na}{i} resonance
doublet ($\lambda\simeq 5890$\,\AA), and the \Ion{Mg}{i} triplet ($\lambda\simeq 5180$\,\AA).
Several magnetic DAZ white dwarfs have also been identified, i.e. those with hydrogen
dominated atmospheres, though their field strengths are typically below 1\,MG
\citep{kawka+vennes11-1,farihietal11-2,zuckermanetal11-1,kawka+vennes14-1,kawkaetal19-1}.
With none of the objects published so far demonstrating fields exceeding 11\,MG, calculations
of metals in ultra-strong magnetic fields have thus far not been essential for the analysis of DZH spectra.

\footnotetext{Note that the thesis of \citet{Hollands-thesis}
identified a further seven low-field magnetic objects in the \citet{hollandsetal17-1} DZ sample,
with field strengths between $250\pm30$\,kG to $510\pm40$\,kG.}

In this work we investigate \WDlong\ (hereafter \WD), a faint
($G$=20.1\,mag) magnetic DZ white dwarf
with a peculiar spectrum with a sufficiently strong magnetic field that
spectral features are almost entirely unrecognisable. In Section~\ref{sec:obs}
we present our observations as well as public data on \WD. In
Section~\ref{sec:atoms} we discuss our finite-field coupled-cluster
calculations for metals in strong magnetic fields. In Section~\ref{sec:lines},
we make use of our atomic data calculations to identify the spectral lines of
\WD\ while simultaneously measuring the strength of its magnetic field. 
In Section~\ref{sec:magmodel}, we attempt to model the field structure of \WD,
while in Section~\ref{sec:discussion} we discuss the applicability of our atomic data
to higher field strengths and use in model atmospheres,
with our conclusions presented in Section~\ref{sec:conc}.

\begin{figure*}
	\includegraphics[width=\textwidth]{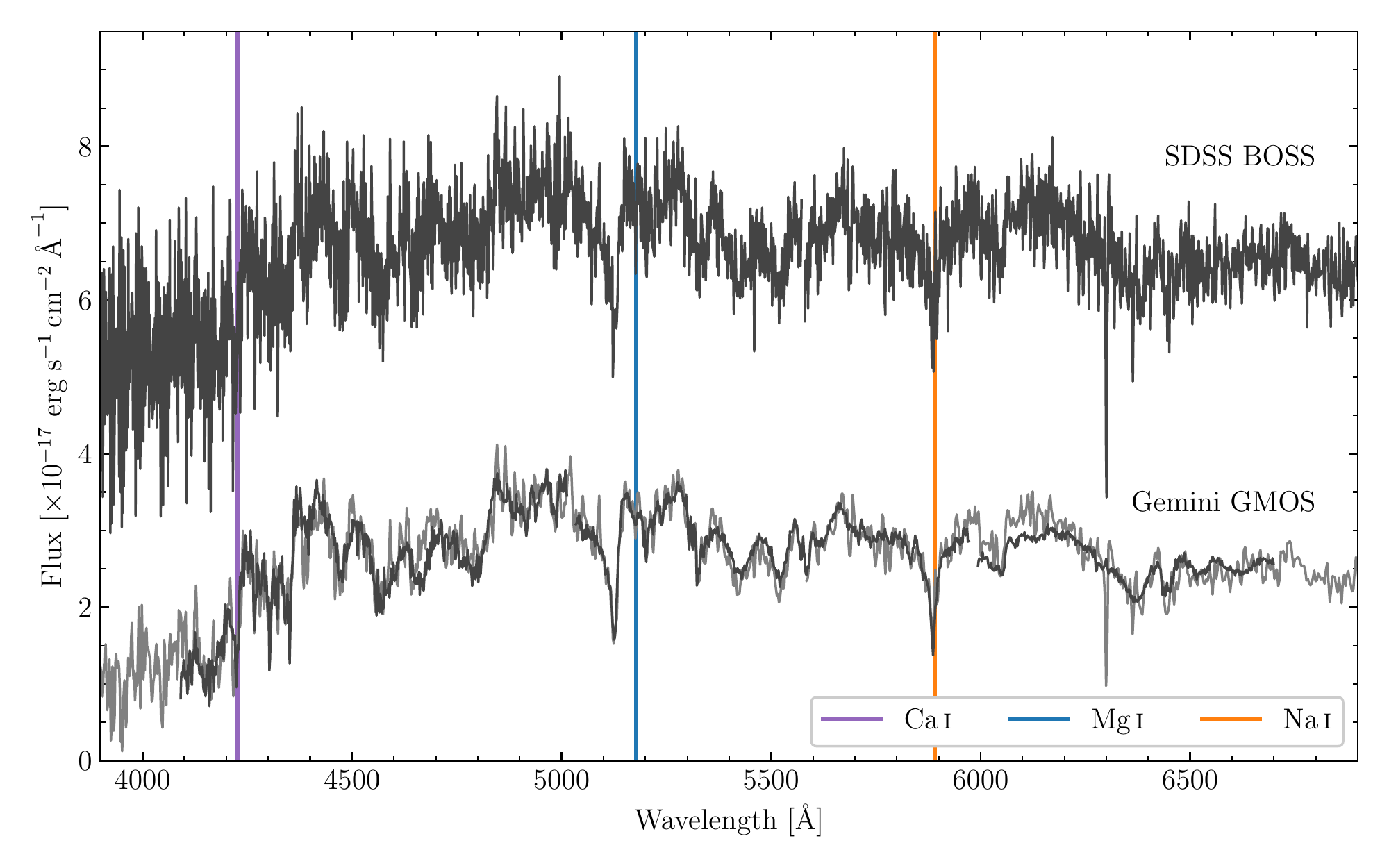}
    \caption{SDSS BOSS and Gemini GMOS spectra of \WD\ ($G$=20.1\,mag). The SDSS
    spectrum is shifted upwards by $4\times10^{-17}$\,\jangstrom. Behind the Gemini
    spectrum, we show the SDSS spectrum again (light grey),
    but convolved to a resolving power of $R=1100$ for direct comparison,
    demonstrating the virtually unchanged spectrum over two years.
    The zero-field air wavelengths of \Ion{Ca}{i}, \Ion{Mg}{i}, and \Ion{Na}{i}
    are shown by the solid vertical lines.
    }
    \label{fig:spec_both}
\end{figure*}

\section{Observations}
\label{sec:obs}

\subsection{SDSS}
\label{sec:sdss}

\WD\ was originally observed in SDSS using the
BOSS spectrograph (Baryon Oscillation Spectroscopic Survey), first published in
SDSS Data Release 12 (plate-MJD-fiberID 7114-56748-0973). The SDSS spectrum is
shown at the top of Figure~\ref{fig:spec_both}. This spectrum was first classified as a
candidate DZH white dwarf by \citet{kepleretal16-1}. This object also appeared
in the DZ sample of \citet{hollandsetal17-1}, where it was suggested to have a
magnetic field exceeding $20$\,MG.

The overall slope of the spectrum appears consistent with a cool white dwarf with
effective temperature (\Teff) in the range $5000$--$7000$\,K, but is otherwise
highly unusual, exhibiting a myriad of unidentified features.
In particular several bands of broad features are seen near 4700\,\AA, 5500\,\AA, and
6400\,\AA. However, two sharper absorption features stand out as resembling atomic lines.
One of these appears at about 5890\,\AA, and so could be from the \Ion{Na}{i}-D
resonance doublet (which in the absence of a magnetic field would appear blended here).
The other sharp feature is located at $\simeq5125$\,\AA, and due to its asymmetry
resembles the \Ion{Mg}{i}-b triplet which is commonly observed in cool DZ white dwarfs
where the asymmetry arises from neutral broadening by
helium atoms in a dense, helium dominated atmosphere
\citep{allardetal16-1,hollandsetal17-1,blouin20-1}. However, while the asymmetry
appears qualitatively similar, the wavelength is bluer by about 50\,\AA{}  than
should be the case for the Mg triplet.
While the SDSS spectrum does extend to $10,400$\,\AA, we see no evidence for other
absorption features beyond what is shown in Figure~\ref{fig:spec_both}.
With none of the spectral features firmly identified, we speculated that \WD\ is a strongly
magnetic DZ white dwarf, where the quadratic Zeeman effect is no longer negligible,
causing additional shifts of Zeeman-split spectral lines,
and resulting in the appearance of many unidentified features in the spectrum.

The SDSS spectrum itself is composed of four sub-spectra, each taken with
900\,s exposure times. While these individual spectra are extremely noisy,
owing to the faintness of \WD, smoothing the data and down-sampling
hinted at possible variability between exposures. Because magnetic white dwarfs are
known to have rotation periods of minutes to days \citep{brinkworthetal13-1,kilicetal21-1},
we considered the possibility of spectral line shapes/positions evolving with
rotational phase. We therefore sought to obtain higher quality spectra of \WD\ in
order to confirm this rotation, as well potentially identify spectral lines.

\subsection{Gemini}
\label{sec:gemini}

We obtained additional spectra using the GMOS (Gemini Multi Object
Spectrograph) instrument on the Gemini North telescope on April 1st 2016 (exactly two years
after the SDSS spectrum was taken). The
instrumental setup used the B600\_G5307 grating with a 0.75\,arcsec slit,
giving us a resolving power of about 1100 at 4600\,\AA. In total we took 17
exposures lasting 628\,s each, separated by 15\,s of readout time. The GMOS
detector uses three CCDs which covered 4100--7000\,\AA\ with our
instrumental setup. This results in two $\simeq25$\,\AA\ gaps between each CCD
with no spectral coverage, though these did not cover any important
features identified from the SDSS spectrum (Figure~\ref{fig:spec_both}).

We reduced the GMOS spectra using the \texttt{starlink} distribution of software
for bias-subtraction, flat-fielding, and optimal-extraction \citep{horne86-1,marsh89-1}
of the spectral trace. Wavelength-calibration was performed using 
\texttt{molly}\footnote{The software \texttt{molly} can be found at
\url{https://cygnus.astro.warwick.ac.uk/phsaap/software/}}.
For flux-calibration, we initially used our observed flux standard, EG\,131, but
found this gave unsatisfactory results, since it was observed at the end of the
night, whereas our science observations were observed at the start.
We instead made use of the SDSS spectrum from Section~\ref{sec:sdss}, as the SDSS
flux calibration are typically accurate to 1\,percent. For each chip we took
the ratio of the spectra (in units of counts) and the already flux-calibrated SDSS
spectrum, re-binned onto the same wavelengths as the GMOS spectra.
We then fitted third-order polynomials to these ratios to define a
calibration function, which we then used to re-scale the Gemini spectra into flux units.
Note that fluxes redwards of 6700\,\AA\ were dominated by telluric absorption and so data beyond
this wavelength were ignored and are not shown in Figure~\ref{fig:spec_both}.

\begin{figure}
	\includegraphics[width=\columnwidth]{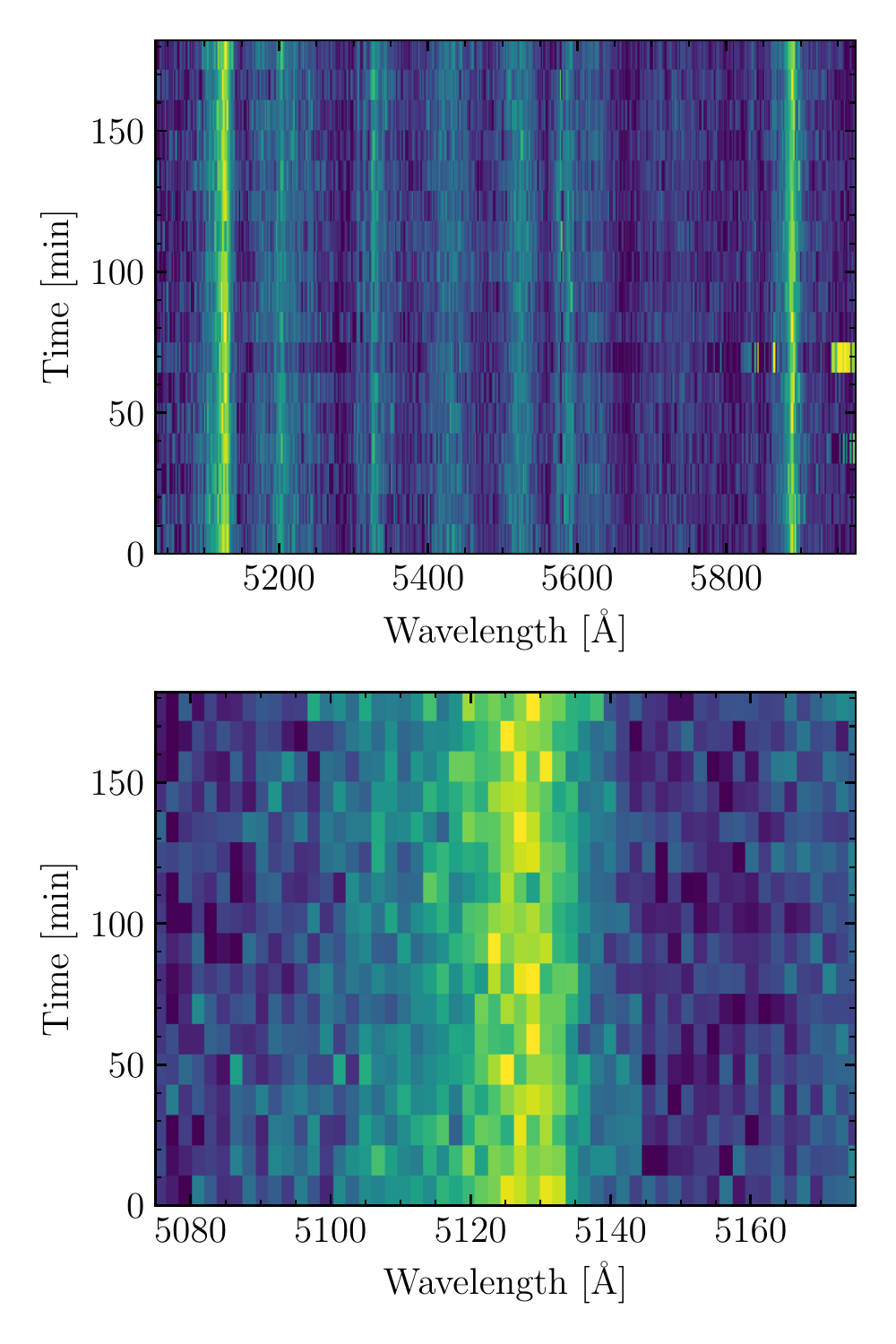}
    \caption{Trailed continuum-normalised spectra for our Gemini observations
    of \WD. The top panel shows the entirety of chip-2, which contains both of
    the sharp features suggested to be from Mg and Na. The bottom panel shows a
    Zoom-in of the suggested Mg line, demonstrating an absence of spectral
    variability on a 3\,hr timescale.}
    \label{fig:trailed}
\end{figure}

Our initial goal for these time-resolved spectra was to search for variability, which
may arise from rotation of a magnetic white dwarf, bringing different parts of the magnetic
field structure into view, and thus causing Zeeman components to change in shape
and wavelength. We show the trailed, normalised Gemini spectra in Figure\,\ref{fig:trailed}
for chip-2 of GMOS. This chip has the largest spectral signal-to-noise ratio (S/N), and
contains many of the unassigned spectral features, including the proposed \Ion{Mg}{i}
and \Ion{Na}{i} lines. In the bottom panel, we show a zoom-in of the suggested \Ion{Mg}{i}
line, which because of the large shift from the rest-wavelength, should be particularly
sensitive to changes in the magnetic field (if it is indeed Mg).
We do not detect variability in any of the spectral features,
suggesting a lack of rotation on time scales of a few hours.

Given the lack of variability between our 17 spectra, we chose to co-add these
into a single high S/N spectrum. We show this in the bottom of Figure~\ref{fig:spec_both}
(dark grey). This is compared with the SDSS spectrum (light grey) which has been convolved
to the same spectral resolution as our Gemini data. Almost all features appear
unchanged, with perhaps only minor differences in the core strengths of the 5400\,\AA\ and
5500\,\AA\ features, and a slight change in wavelength of the feature at 4650\,\AA.
This comparison demonstrates a lack of variability on a time scale of two years.

With the higher S/N spectrum, the proposed \Ion{Na}{i} and \Ion{Mg}{i} lines are
seen to be blue shifted by 5.5\,\AA\ and 52\,\AA\ respectively. The asymmetric nature
of the latter (discussed in Section~\ref{sec:sdss}), is also much clearer.
For the proposed \Ion{Na}{i}
line this could be plausibly explained as a $\simeq 300$\kms\ blue shift (not including
any gravitational redshift from the white dwarf) \emph{if} \WD\ is a halo object. That being
said, the much slower $18\pm2$\,\kms\ tangential velocity from \emph{Gaia} EDR3
(see Section \ref{sec:gaia}) argues against this explanation. Furthermore,
such an explanation is effectively ruled out by the proposed \Ion{Mg}{i} line,
since its much larger wavelength shift would correspond to a velocity shift of about 3000\,\kms.
Therefore magnetism remains a more likely hypothesis for explaining the spectrum
of \WD. In addition to the lines observed from the SDSS spectrum, the Gemini spectrum
also reveals the possible presence of the \Ion{Ca}{i} resonance line
(Figure~\ref{fig:spec_both}, purple), with a small blue shift of 1.6\,\AA.

\subsection{Gaia}

\label{sec:gaia}

Despite its curious spectrum containing many anomalous features precluding
obvious spectroscopic classification, the measured non-zero proper-motion by
SDSS confirms that \WD\ is a galactic object. However, without knowing the
absolute brightness of this star, \WD\ could not be claimed to be a white
dwarf with certainty.

\begin{figure}
	\includegraphics[width=\columnwidth]{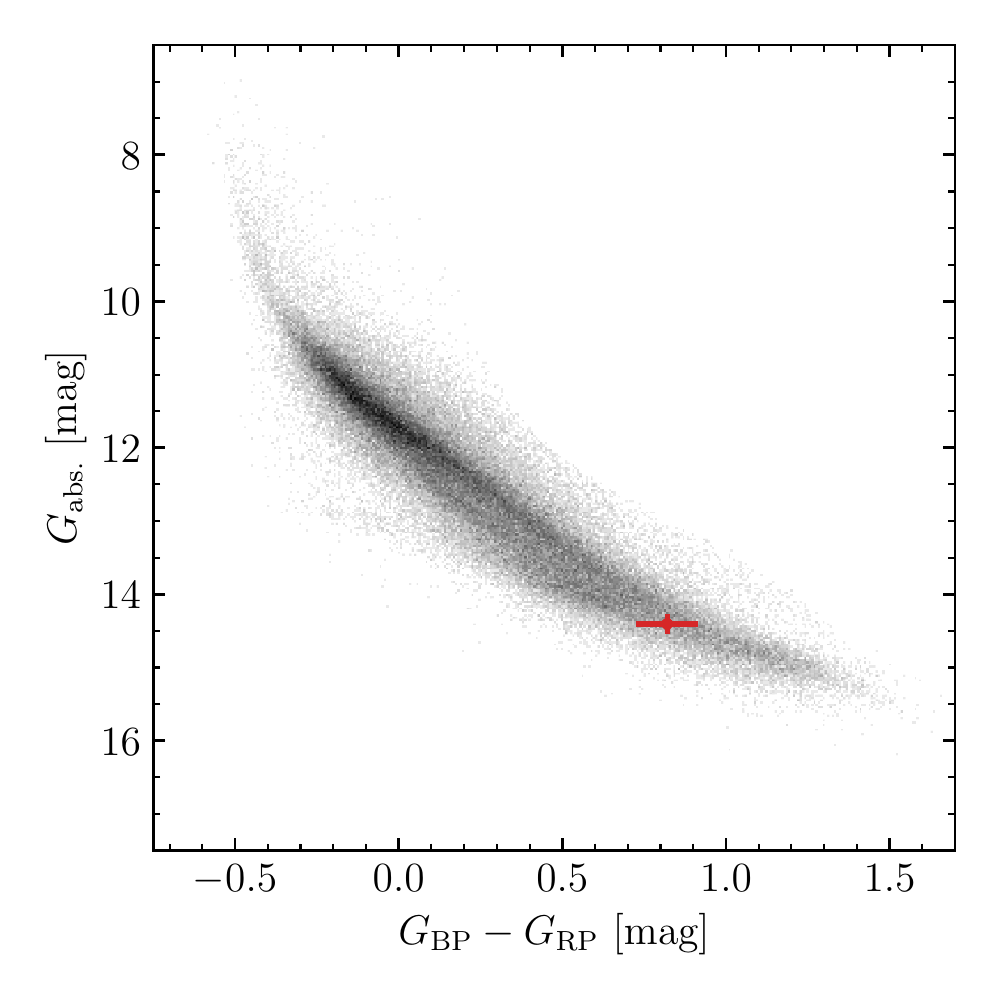}
    \caption{Gaia EDR3 Hertzsprung-Russel diagram showing the location of \WD\ (red)
    compared with the white dwarf cooling sequence (grey histogram). The error
    bars represent $1\sigma$ uncertainties.}
    \label{fig:HRD}
\end{figure}

In April 2018, the second data release (DR2) from \emph{Gaia} space mission
made public approximately 1 billion parallaxes \citep{gaiaDR2-collab-1}. This
included \WD\ which had a measured parallax of $7.79\pm0.68$\,mas, confirming
this stars location along the white dwarf cooling track within the
Hertzsprung-Russel diagram (HR-diagram). In December 2020 a refined parallax of
$7.24\pm0.46$\, mas was made available from \emph{Gaia} EDR3 \citep[early data
release 3,][]{gaiaEDR3-collab-1} corresponding to a distance of $138.8\pm9.0$\,pc. The EDR3
HR-diagram is shown in Figure~\ref{fig:HRD}. \WD\ is indicated by the red
point, and is compared against a background of white dwarfs selected from
\citet{gentilefusilloetal21-2} with $\mbox{\texttt{PWD}} > 0.75$ and
$\mbox{\texttt{parallax\_over\_error}} > 20$. From its location in the
HR-diagram, it is clear that \WD\ is a cool white dwarf with a typical mass.
Therefore \citet{gentilefusilloetal21-2} found $\Teff=5810\pm460$\,K and $\log
g =8.17\pm0.33$ fitting the \emph{Gaia} photometry with pure hydrogen
atmosphere models, and $\Teff=5680\pm470$\,K and $\log g = 8.08\pm0.33$ for a
pure helium atmosphere models. Interestingly, if Figure~\ref{fig:HRD} is
recreated using \emph{Gaia} DR2 data, \WD\ appears to be offset from the white
warf sequence towards higher masses, with \citet{gentilefusilloetal19-1}
finding $\Teff=6990\pm710$\,K and $\log g = 8.73\pm0.29$ for hydrogen
atmosphere models, and $\Teff=6870\pm750$\,K and $\log g = 8.67\pm0.33$ for
helium atmosphere models. That said, these parameter shifts amount to only
$1.4\sigma$ changes at most and so are in statistical agreement.

\section{Atomic data calculations}
\label{sec:atoms}
To test our hypothesis that \WD\ is a highly magnetic DZ white dwarf, we required
accurate wavelengths of (at the very least) the \Ion{Na}{i} and \Ion{Mg}{i}
lines as a function of the magnetic field. For large magnetic field strengths, however, approaches that are based on a perturbative treatment of the magnetic field are no longer adequate and hence accurate finite-field quantum-chemical methods need to be employed. In these methods, the magnetic field is treated explicitly in the calculation of ground-state energies, excitation energies, and transition-dipole moments, thereby employing the electronic Hamiltonian for an $N$-electron system in a static magnetic field in the $z$-direction (the gauge-origin is here in the origin of the coordinate system)
\begin{equation}
\hat H = \hat H_0 + \frac{1}{2}  B \hat L_z +  B  \hat S_z + \frac{1}{8}B^2\sum_i^N (x_i^2+y_i^2),
\end{equation}
where $B$ is the magnetic-field strength and  $\hat H_0$ is the field-free atomistic (or molecular) Hamiltonian, consisting of the kinetic energy of the electrons, the nuclear-electronic potential and the electron-electron repulsion. $\hat L_z= \sum_i^N \hat l_{i,z}$ and $\hat S_z=\sum_i^N \hat s_{i,z}$ are the $z$-components of the angular momentum operator, and spin, respectively. The terms linear in the magnetic field are the orbital-Zeeman (responsible for the splitting of the orbitals) and spin-Zeeman terms (responsible for splitting according to spin parallel or antiparallel to the magnetic field), respectively. The quadratic term is referred to as diamagnetic contribution which always increases the energy of the system.
As in the field-free case in quantum chemistry, FCI theory is not applicable for problems like ours due to its high computational cost. Instead, Coupled-Cluster (CC) theory \citep{ccbook} can be used, which has a more economical computational scaling. 
CC methods work with an exponential parametrization of the wave function 
$\Psi_\mathrm{CC}=\mathrm{e}^{\hat T} \Phi_0$, where $\hat T=\hat T_1 + \hat T_2 + \dots + \hat T_N$ is the so-called cluster operator generating excitations. $\hat T$ contains amplitudes (weighting coefficients in the wave functions) that are determined by solving the CC equations 
\begin{equation}
\langle \Phi_I\mid \mathrm{e}^{-\hat T}\hat H \mathrm{e}^{\hat T}\mid \Phi_0\rangle =0.
\end{equation}
The CC energy is then given as 
\begin{equation}
E_\mathrm{CC} = \langle \Phi_0\mid \mathrm{e}^{-\hat T}\hat H \mathrm{e}^{\hat T}\mid \Phi_0\rangle.
\end{equation}
Truncations in $\hat T$ as well as limiting the projection space define approximate CC schemes. 
For example, CC `singles doubles' (CCSD) is defined with 
\begin{equation*}
\hat T^{\mathrm{CCSD}}=\hat T_1+ \hat T_2
\end{equation*} 
and projection on singly and doubly excited determinants. Analogously, in CC `singles doubles triples' (CCSDT), $\hat T$ is truncated to 
\begin{equation*}
\hat T^{\mathrm{CCSDT}} = \hat T_1+ \hat T_2 +\hat T_3
\end{equation*}
and projection is additionally also performed on triply excited determinants. 
While CC is used to describe the ground-state wave function, Equation-of-Motion-CC (EOM-CC) \citep{ccbook} can also describe electronically excited states (EE). An operator $\hat R$, parametrized similarly as $\hat T$ acts on a CC wave function $\Psi_\mathrm{exc} 
= \hat R 
\Psi_\mathrm{CC} $. The corresponding amplitudes are determined via the solution of the eigenvalue problem in matrix form 
\begin{equation}
\mathbf{\bar H r} = \Delta \mathbf{E}_\mathrm{exc}\mathbf{r}
\end{equation} 
in which an element of the matrix $\mathbf{\bar H}$ is given as 
\begin{equation}
\bar H_{IJ} = \langle \Phi_I \mid
\mathrm{e}^{-\hat T}  (\hat H  -  E_\mathrm{CC} )  \mathrm{e}^{\hat T}
\mid \Phi_J \rangle
\end{equation}
and the vector $\mathbf{r}$ contains the amplitudes for the excitations.
An overview of ff-CC and ff-EOM-CC methods can be found in \citet{perspectiveCC}. 
In this work, we have used various flavors of ff-CC theory \citep{Stopkowicz2015, petrosnew} and ff-EOM CC theory, implemented within the \citetalias{QCUMBRE} program package \citep{Hampe2017},
 to determine excited states and hence transition wavelengths \citep{Hampe2017,triplespaper, petrosnew}. The underlying calculation of the  reference $| \Phi_0\rangle$ is performed with the \citetalias{cfour} program package \citep{cfourpaper}. In the EOM-framework, we have employed the methods for electronic excitations (EE), spin flip (SF), adding electrons (EA, electron attachment) and removal of electrons (IP, ionization potential). Oscillator strengths are also treated at the expectation value ff-EOM-CC level \citep{Hampe2019} which enables the prediction of field-dependent intensities.
The transitions for which we have performed ff-calculations are displayed in Table~\ref{tab:transitions}.
\begin{table}
    \centering
    \caption{\label{tab:transitions} Level information for the transitions we have
    performed ff-calculations for. Wavelengths (air) correspond to field-free transitions,
    which in the case of multiplets corresponds to the average wavelength given in the
    NIST database (weighted by oscillator strength).
    }
    \begin{tabular}{lcll}
        \hline
        Ion & Wavelength [\AA] & Lower state & Upper state \\
        \hline
        \Ion{Na}{i}  & 5892 & $^2S_g ([\text{Ne}]3s)  $ & $^2P_u ([\text{Ne}]3p)$   \\
        \Ion{Mg}{i}  & 5178 & $^3P_u ([\text{Ne}]3s3p)$ & $^3S_g ([\text{Ne}]3s4s$) \\
        \Ion{Ca}{i}  & 4227 & $^1S_g ([\text{Ar}]4s^2)$ & $^1P_u ([\text{Ar}]4s4p)$ \\
        \Ion{Ca}{i}  & 6142 & $^3P_u ([\text{Ar}]4s4p)$ & $^3S_g ([\text{Ar}]4s5s)$ \\
        \Ion{Ca}{ii} & 3945 & $^2S_g ([\text{Ar}]4s)  $ & $^2P_u ([\text{Ar}] 4p)$  \\
        \hline
    \end{tabular}
\end{table}
The data for Na has partly already been available in \citet{triplespaper}. The latter work is also the basis for the computational protocol. We will here only mention the most important points and refer to \citet{triplespaper} for further details. 
For all transitions, the calculations were performed for magnetic fields ranging between 0.00--0.04\,B$_0$, with the atomic unit of the magnetic field, B$_0$ $\simeq 2350.518$\,MG, using a 0.004\,B$_0$ step and between 0.04--0.20\,B$_0$ using a 0.02\,B$_0$ step.
In the protocol, a corrected excitation energy is computed according to 
\begin{equation}
    \Delta E_{\mathrm{exc}}^\mathrm{corrected} = \Delta E_\mathrm{exc} + \Delta E_\mathrm{basis} + \Delta E_\mathrm{triples},
\end{equation}
where $\Delta E_\mathrm{exc}$ is the excitation energy computed using a large uncontracted augmented one-electron basis set. $\Delta E_\mathrm{basis}$ is a term correcting the one-electron basis-set error as described in \citet{basiscorrection} by extrapolating a basis-set limit $E^\infty$
 based on uncontracted basis sets of the type aug-cc-pCVXZ \citep{Dunningaug1, DunningPC}, abbreviated as aCXZ, where X is the cardinal number. It is given as $\Delta E_\mathrm{basis}= \Delta E^\infty - \Delta E_\mathrm{exc}$ with 
 \begin{equation}
 \label{Eextrap}
 \Delta E^\infty =  \frac{\Delta E_\mathrm{exc}^\mathrm{aCXZ}X^3-\Delta E_\mathrm{exc}^{\mathrm{aCYZ}}Y^3}{X^3-Y^3}.
\end{equation}
 The $\Delta E_\mathrm{triples}= E_\mathrm{triples}^\mathrm{aCXZ}- E_\mathrm{exc}^\mathrm{aCXZ}$ correction accounts for the error which stems from truncating the CC expansion and involves computations at the ff-EOM-CCSDT \citep{triplespaper}, ff-EOM-CC3 \citep{petrosnew} or ff-EOM-CCSD(T)(a)* \citep{PertTriples,petrosnew2} levels of theory for $E_\mathrm{triples}^\mathrm{aCXZ}$ using a smaller basis set. 
 The accuracy and cost is typically CCSDT ($O(M^8)$)> CC3 ($O(M^7)$)> CCSD(T)(a)* ($O(M^7)$) where $M$ is the number of basis functions. 
 In the latter two, triple-excitations are treated in a perturbative manner. CC3 is iterative while CCSD(T)(a)* is not. The latter is a very good and relatively cheap option when the target-states are characterised mostly by single-excitation character. 
 The dimensionless oscillator strengths $f_{IJ}$ were calculated according to 
\begin{equation}
    \label{oscillator}
    f_{IJ} = \frac{2}{3}(\Delta E_{IJ}) |\mu_{IJ}|^2,
\end{equation}
where $\Delta E_{IJ}$ is the excitation energy from states $I$ to $J$ and $\mu_{IJ}$ is the corresponding transition-dipole moment,
and where both $\Delta E_{IJ}$ and $\mu_{IJ}$ are in atomic units.
After converting the (field-dependent) excitation energies to transition wavelengths, the resulting $B-\lambda$ curves were shifted to start at the zero-field values taken from the NIST database \citep{NIST_ASD} thereby correcting for remaining errors of our predictions at zero field. The spin-orbit contributions have been averaged out as their contribution is expected to be small for stronger fields. By the shift made to the NIST data, field-free scalar-relativistic effects are implicitly accounted for.
For the time being, we are neglecting relativistic effects and in particular their dependence on the magnetic field in our calculations as the effects are expected to be small for strong magnetic fields. This approximation is better for the lighter elements Na and Mg than for the heavier Ca.
The specific details on the calculations are collected in Table~\ref{tab:calcs}.
\begin{table*}
    \centering
    \caption{\label{tab:calcs} 
    Detailed information on ff-EOM calculations for the respective transitions. If not specified otherwise, $\Delta E_{IJ}$, see Eq. \eqref{oscillator}, has been calculated at the same level as $\mu_{IJ}$
    }
    \begin{tabular}{lcllll}
        \hline
        Transition & Basis functions & $\Delta E_\mathrm{exc}$ & $\Delta E_\mathrm{basis}$ & $\Delta E_\mathrm{triples}$ & $\mu_{IJ}$ \\
        \hline 
        \Ion{Na}{i} & Cartesian & EE-CCSD/aCQZ & EE-CCSD/aCXZ, X=T, Q & CCSDT/aCTZ & EE-CCSD/aCQZ \\
        \Ion{Mg}{i} & Spherical & EE-CCSD/aC5Z & EE-CCSD/aCXZ, X=Q, 5 & CC3/aCQZ & EE-CCSD/aC5Z \\ 
        \Ion{Ca}{i} 4227 & Spherical & EE-CCSD/aC5Z & EE-CCSD/aCXZ, X=Q, 5 & EE-CC3/aCQZ & EE-CCSD/aCQZ$^{(a)}$ \\
        \Ion{Ca}{i} 6142 & Spherical & SF-CCSD(T)(a)*/aC5Z & SF-CCSD(T)(a)*/aCXZ, X=Q, 5 & No further triples correction & SF-CCSD/aC5Z$^{(b)}$ \\
        \Ion{Ca}{ii} & Spherical & EA-CCSD/aC5Z & EA-CCSD/aCXZ, X=Q, 5 & EE-CCSD(T)(a)*/aCQZ & EA-CCSD/aC5Z$^{(c)}$ \\
        \hline
    \end{tabular}
    Notes: ($a$) $E_{IJ}$ calculated using EE-CC3, ($b$) Reference for SF calculations: $^1S_g$ ([Ar] $4s^2$), ($c$) Reference for EA calculations: $^1S_g$([Ar])
\end{table*}

\begin{figure*}
	\includegraphics[width=\textwidth]{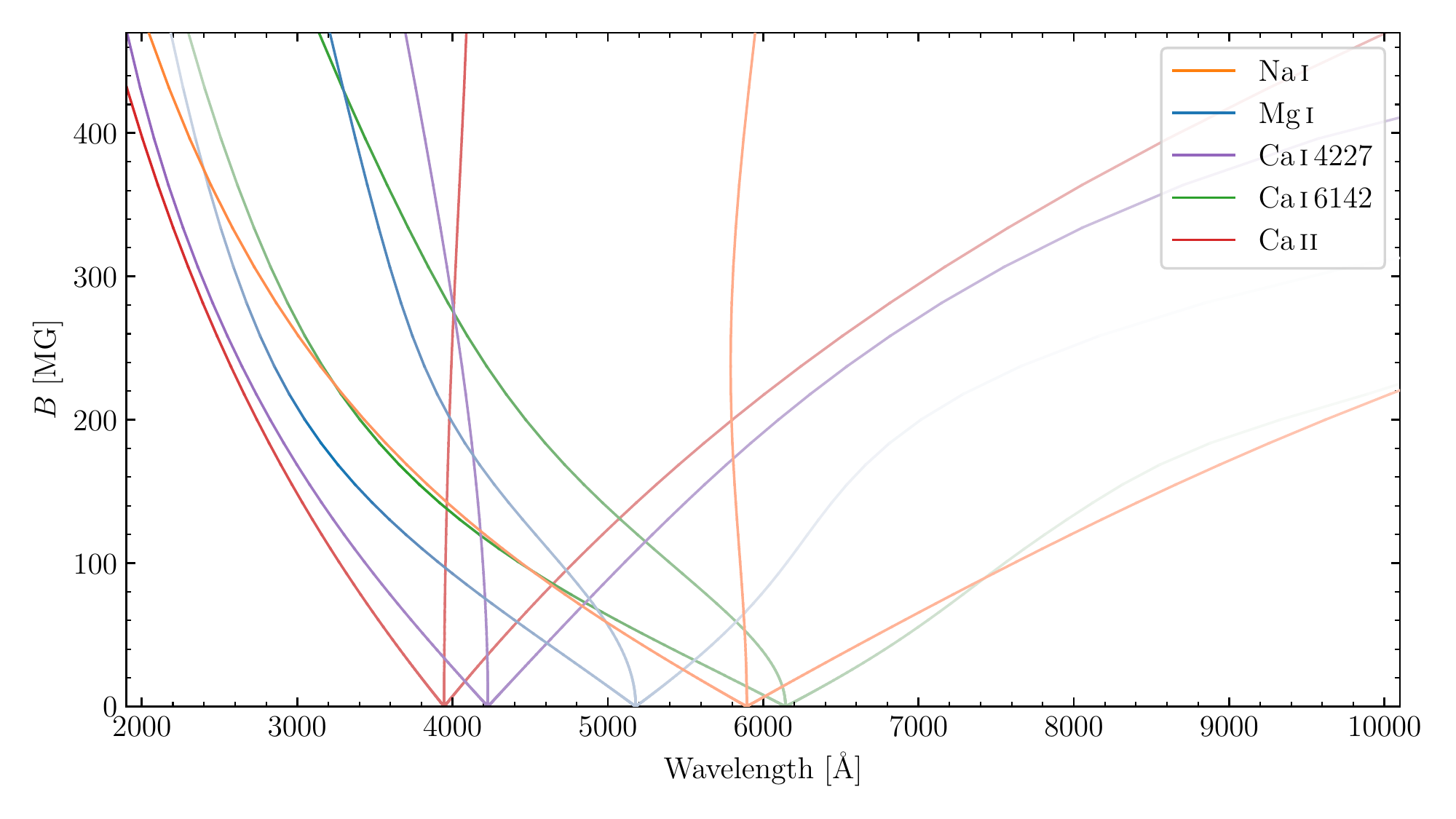}
    \caption{Calculated transition wavelengths as a function of field strength.
    For each Zeeman triplet, the line opacities are scaled to the oscillator
    strengths.}
    \label{fig:spaghetti}
\end{figure*}

The predicted transition wavelengths and oscillator strengths can be found in Tables~\ref{tab:data_NaI}--\ref{tab:data_CaI6142}. Additionally, the obtained $B-\lambda$ curves are shown in Fig. \ref{fig:spaghetti}. The intensity of the transitions, i.e. oscillator strengths, are indicated via the opacities of the curves. As all of the investigated transitions are of $ns \rightarrow np$ or $np \rightarrow (n+1)s$ type, where $n$ is the main quantum number of the orbital (without field), there is in all cases a splitting into three components, i.e., the central $\pi$ (transition from/into a $p_0$ orbital) as well as the two $\sigma$ (transition from/into $p_{+1}$ and $p_{-1}$) components\footnote{Note that in the magnetic field, the SO(3) symmetry is lowered to $C_{\infty h}$ but we will, for simplicity, still refer to field-free state and orbital classifications.}. As can be seen here, the splitting is only linear for fields below about 5--10\,MG while for higher field strengths, the form of the $B-\lambda$ curves becomes much more complicated. The distortion from a simple Zeeman behaviour is transition dependent: 
For the $np\rightarrow (n+1)s$ transitions (Mg and \Ion{Ca}{i}~6142), the influence of the magnetic field on the central $\pi$ component is much more pronounced than for the $ns \rightarrow np$ transitions (Na, \Ion{Ca}{ii}, \Ion{Ca}{i}~4227). The principal reason for this behaviour is that in the former case the transitions are between  orbitals of different main quantum numbers. The orbitals hence experience a different amount of polarisation through the magnetic field, i.e. those of higher main quantum number are polarised more strongly due to their more diffuse nature.  Effectively this means that the $s$ and $p_0$ orbitals and the respective electronic states, don't evolve in a parallel manner with increasing magnetic field.
Hence, in contrast to the simple perturbative picture, the central $\pi$ component is no longer constant with increasing magnetic field strength. In addition, the transitions with decreasing energy difference in the magnetic field, i.e., $ns\rightarrow np_{-1}$ and $np_{+1}\rightarrow (n+1)s$ become less relevant for observations, as they decrease in intensity (see Equation~\eqref{oscillator}). In addition, small changes in the magnetic field lead to large changes in the transition wavelength and hence such transitions will be blurred out in the spectra for strong fields. 
A more detailed discussion on the form of the energy levels and the resulting for of the $B-\lambda$ curve of the Mg transition can be found in \citet{petrosnew}. As noted in \citet{triplespaper}, high-accuracy predictions are required as even the prediction for the transition least affected by the magnetic field, i.e., the central $\pi$ component of Na can vary by up to 100\,\AA\ depending on the level of theory and basis set used.\footnote{Note that the uncertainty of the predicted transition wavelengths is not only dependent on the accuracy of the method but also on the position of the absorption peak. }
\section{Line identification}
\label{sec:lines}

With the wavelengths and oscillator strengths calculated in Section~\ref{sec:atoms},
we were able to compare these with the spectrum of \WD. With no immediate indication
of which spectral features could correspond to the
$\sigma$-components of the calculated transitions, we began by restricting ourselves
to the $\pi$-components only. In Section~\ref{sec:obs} we identified possible
$\pi$-components of \Ion{Na}{i}, \Ion{Mg}{i}, and \Ion{Ca}{i} in the SDSS and GMOS spectra,
based on the sharpness of the lines, rough proximity in wavelengths to the field-free values,
and characteristic asymmetry in the case of Mg.

\begin{figure*}
	\includegraphics[width=\textwidth]{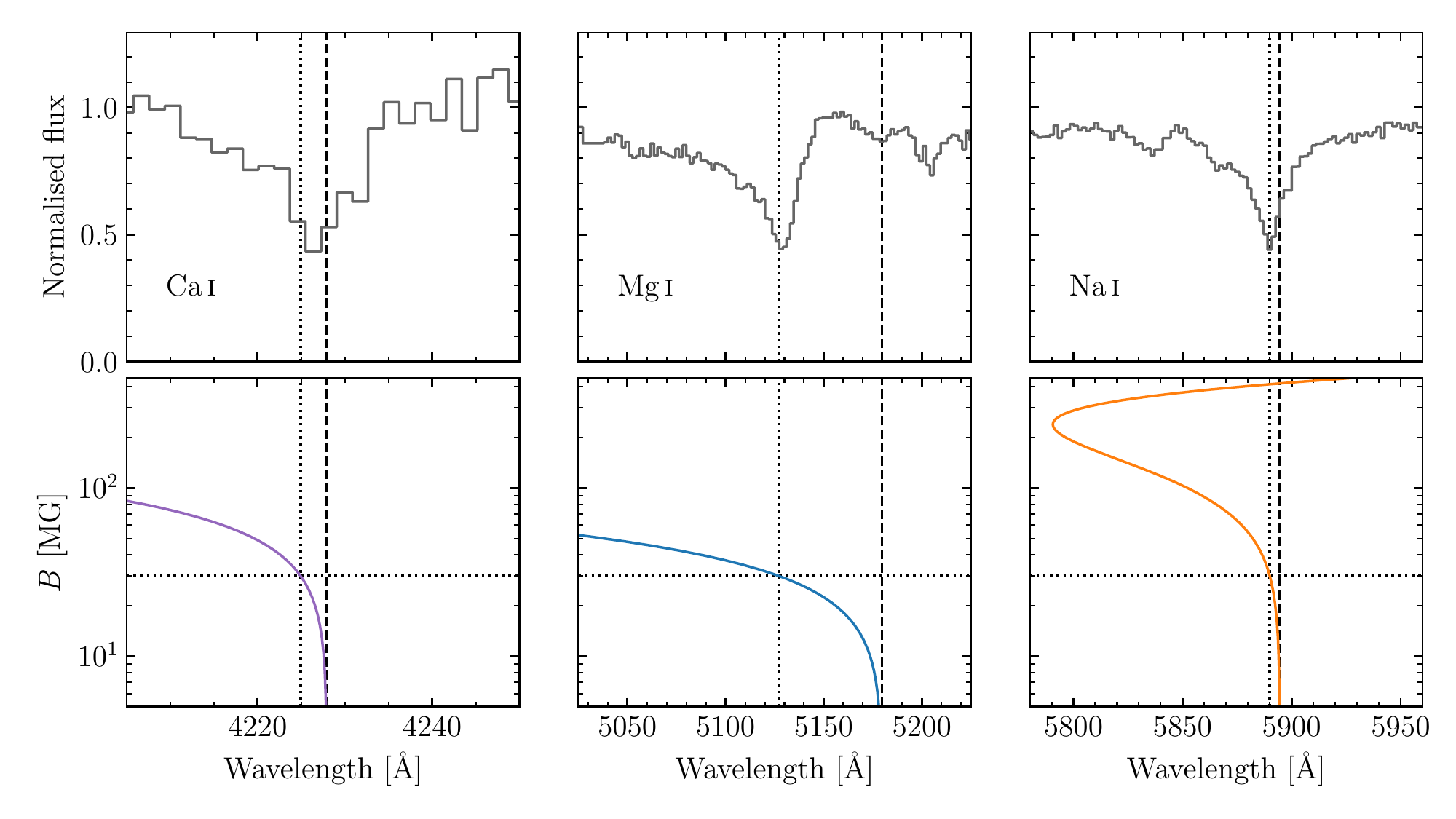}
    \caption{Top row: Spectral regions covering the suspected $\pi$-components
    of \Ion{Ca}{i}, \Ion{Mg}{i}, and \Ion{Na}{i}. Bottom row: Predicted wavelengths
    for the corresponding $\pi$-components as a function of field strength. In all panels
    the black dashed lines indicate the field-free vacuum wavelengths for each line,
    whereas the dotted lines indicate the wavelengths expected for a 30\,MG field.
    }
    \label{fig:pi_zooms}
\end{figure*}

We compare these lines to our calculated
wavelengths as a function of field strength in the top panels of Figure~\ref{fig:pi_zooms}.
From the bottom-right panel, it is clear that the Na line shift could be explained by
either a relatively small field of $\simeq30$\,MG or much larger field of $\simeq410$\,MG,
owing to a turnaround in wavelength at $\simeq240$\,MG.
This degeneracy is entirely resolved by the large shift of the Mg line which has
only one wavelength solution and is also consistent with a field of $\simeq 30$\,MG.
Thus, to our surprise, the peculiar spectrum of \WD\ (Figure~\ref{fig:spec_both})
can not result from a field in the regime of 100s of MG, but is best explained
by a field strength an order of magnitude lower,
though notably still a factor three higher than all previously identified DZH white dwarfs 
\citep{hollandsetal15-1,hollandsetal17-1,dufouretal15-1}.

For \Ion{Ca}{i} the match in wavelength is quite poor, though thus far we have neglected
wavelength shifts that may arise from radial motion and gravitational redshift, the latter
of which could be on the order of 100\,\kms\ if \WD\ is particularly massive,
which is typically the case for magnetic white dwarfs \citep{liebert88-1,kawkaetal20-1,ferrarioetal20-1}.
Additionally, the absent treatment of relativistic effects may here
play a role in the quality of the prediction.
It is also clear that at 30\,MG, the predicted wavelength for Mg is a similar amount
bluer than the line centre (though with greater relative accuracy).
To account for this we fitted the field strength and radial velocity simultaneously.
We measured the line centres for all three $\pi$-components by simply fitting parabolas
to the central few pixels (five for Ca and seven for Mg and Na), constraining them
with uncertainties of 0.1--0.3\,\AA. Performing a least squares fit to the three
line centres, we found a magnetic field strength of $29.92\pm0.05$\,MG
and a redshift of $74\pm8$\,\kms. With these best fitting values the
residuals are $-0.7$\,\AA, $0.0$\,\AA, and $1.8$\,\AA\ for the Ca, Mg, and Na lines, respectively.
This was a clear improvement for \Ion{Ca}{i} and \Ion{Mg}{i}, though provides a somewhat
worse result for the \Ion{Na}{i} line.

\begin{figure*}
	\includegraphics[width=\textwidth]{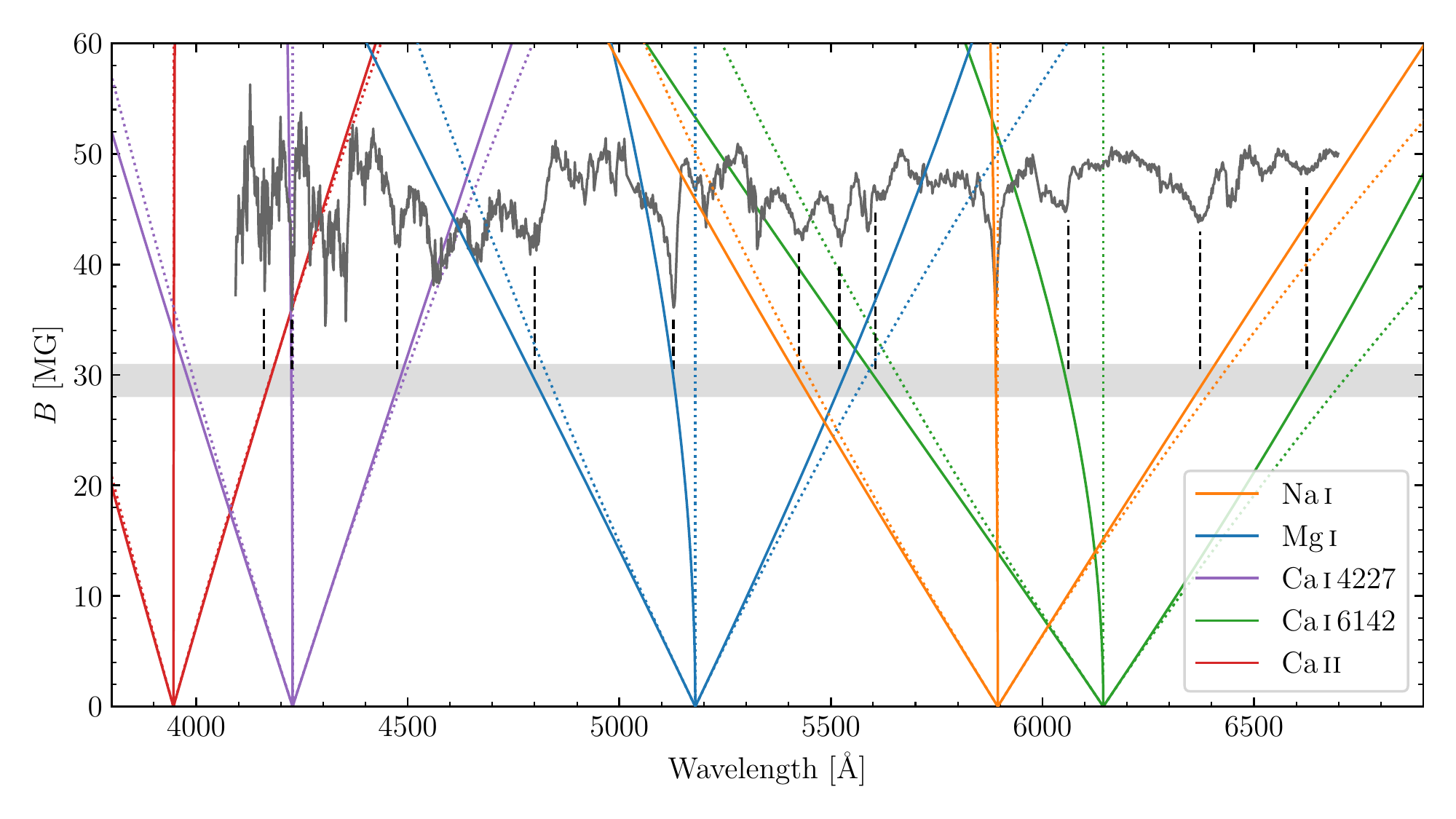}
    \caption{Line identification diagram for \WD. The Zeeman triplets from our
    finite-field coupled-cluster calculations are shown by the solid curves,
    with the na\"ive wavelengths from the linear Zeeman effect indicated by the
    dotted lines. These are plotted over the spectrum of \WD\ (grey), where
    black dashed lines match Zeeman components to features in the spectrum for
    a field strength of approximately 30\,MG (light grey horizontal band).
    }
    \label{fig:lineID}
\end{figure*}

With the field strength established from the $\pi$-components, we could then determine
the expected wavelengths of the $\sigma$-components. We make this comparison in
Figure~\ref{fig:lineID}. We first investigated the components of Na and Mg, with
their $\sigma$-components identified with relative ease. In particular the large
broad feature at $\simeq 6350$\,\AA\ is established as the $\sigma^+$ component of Na, 
which does not appear blended with any of the other nearby features. Near 5500\,\AA\ both
the Na $\sigma^-$ and Mg $\sigma^+$ components are observed, though notably the order
of their wavelengths has swapped due to the components crossing at a field strength
of $\simeq 25$\,MG. The Mg $\sigma^-$ component is inferred to be the broad, asymmetric
feature at $\simeq 4800$\,\AA. The asymmetry appears more extreme than for the
$\pi$-component, which itself is more asymmetric than the $\sigma^+$ component. This
may imply that the degree of neutral broadening affects each component differently,
which perhaps is not surprising given that both the perturbations from neutral
helium atoms and the magnetic field both alter the energy levels of Mg.

Having identified all components from \Ion{Na}{i} and \Ion{Mg}{i}, we proceeded
with classifying transitions from Ca. For the \Ion{Ca}{i} resonance line, we
had already identified the $\pi$-component (rest wavelength at 4227\,\AA; see
Figure~\ref{fig:pi_zooms}, left). As our Gemini GMOS spectrum does not go bluer
than about 4090\,\AA, the $\sigma^{-}$-component is not covered, and so we were
only able to search for the $\sigma^{+}$ component which, at 30\,MG, has an 
expected wavelength of 4475\,\AA. Indeed, a spectral feature was found at this
wavelength which we attribute to the $\sigma^{+}$ component (Figure~\ref{fig:lineID}).

The final Ca transitions are less certain,
though we still make some attempt at their classification. 
For the \Ion{Ca}{ii} Zeeman triplet (H+K resonance doublet in the absence of an
external magnetic field), only the $\sigma^{+}$ component is expected to be
covered by our GMOS spectrum at a field strength of 30\,MG. While we detect
a feature at the expected wavelength of 4160\,\AA\ (Figure~\ref{fig:lineID}),
the signal-to-noise ratio is somewhat poor at this end of the spectrum,
making this assignment less secure. However, it is worth noting that for
a \Teff\ between 5000\,K and 7000\,K, both \Ion{Ca}{i} and \Ion{Ca}{ii} resonance
lines are typically observed together in non-magnetic DZs \citep{hollandsetal17-1}.

Finally we consider the \Ion{Ca}{i} $4p\rightarrow 5s$ transition, which in
the absence of an external magnetic field appears as a triplet (due to the spin-orbit interaction)
centred on 6142\,\AA.
In the presence of a strong magnetic field, this transition appears as a Zeeman triplet exhibiting the strongest quadratic shift of all the transitions
calculated in Section~\ref{sec:atoms}. Nevertheless, weak transitions are observed
at all of the expected wavelengths. Whether this assignment is correct is debatable:
the identified central component at around 6060\,\AA\ shows some asymmetry, as is
observed in the field-free case \citep[see SDSS\,J0916$+$2540 in Figure~10 of ][]{hollandsetal18-1}.
On the other hand, the 6142\,\AA\ triplet is typically much weaker than the \Ion{Ca}{i} 4227\,\AA\
resonance line, and is only usually visible for extremely large calcium abundances.
Yet, in the case of \WD, the established components of the 4227\,\AA\ \Ion{Ca}{i} Zeeman
triplet are not particularly strong, suggesting that the 6142\,\AA\ components would
likely be too weak to be visible. Given the sheer number of unknown features in the spectrum 
of \WD, it is probable that our assignments to the 6142\,\AA\ triplet in
Figure~\ref{fig:lineID} might also originate from some other source.

Many anomalous features in the spectrum of \WD\  remain unclassified.
In particular two strong and broad features are observed
at wavelengths of $\simeq 4570$\,\AA\ and $\simeq 4660$\,\AA, between the
$\sigma^{+}$-component of the \Ion{Ca}{i} resonance line, and the $\sigma^{-}$-component
of \Ion{Mg}{i}. The strength of these features suggest they originate from another
element commonly observed in DZ spectra. With the strongest Na, Mg and Ca lines already
accounted for, the most likely candidate is therefore Fe. In the field-free case,
a large number of \Ion{Fe}{i} lines can be found between 4000--4500\,\AA\
\citep[see][Figure 7]{hollandsetal18-1}. Among the strongest transitions in this
range are the $^{3}F \rightarrow\ ^{5}G$ and $^{3}F \rightarrow\ ^{3}G$ multiplets,
which share the same lower level.
We therefore suggest that the unidentified features
at $\simeq 4570$\,\AA\ and $\simeq 4660$\,\AA\ arise from these iron transitions.
Additional unidentified features include broad absorption around $4300$\,\AA\ 
(between the $\pi$- and $\sigma^{+}$-components of \Ion{Ca}{i}), 
sharp features at $\simeq 5200/5330/5580$\,\AA, and several other features
at $\simeq 6030/6450/6530/6620$\,\AA\ (some of which we were unable to conclusively
assign to the \Ion{Ca}{i} $6142$\,\AA\ multiplet).
We note that the feature near 5200\,\AA\ is close to the field-free wavelength
of the \Ion{Cr}{i} $4s\rightarrow 4p$ triplet (5208\,\AA, vacuum), and
so that feature could plausibly correspond to the $\pi$-component of the
\Ion{Cr}{i} transition. Firmly establishing the origin of these remaining
features necessarily will require additional finite-field coupled-cluster
calculations in the future,
with the above Fe and Cr transitions as the highest priority. For these systems, treatment of field-dependent relativistic effects and a robust treatment of multi-reference character in the electronic structure will be important.

\section{Magnetic field modelling}
\label{sec:magmodel}

With several of the spectral features of \WD\ identified, we finally sought to model
the magnetic field distribution across its surface. For a purely dipolar
magnetic field, the field strength spans a factor of two between the equator and
poles. This results in broadened spectral lines, particularly the
$\sigma$-components due to their stronger wavelength dependence of the field
strength. It is clear from the width of the \Ion{Na}{i} $\sigma^+$ component
that the range of magnetic field strengths on the visible hemisphere of \WD\ spans a
much narrower field range, with Figure~\ref{fig:lineID} suggesting about
24--31\,MG. Thus it is necessary to invoke a field structure more complex than a
centred dipole.

\subsection{The offset dipole model}
\label{sec:offsetdipole}

We chose to use the offset-dipole model, which has been commonly used in the
analysis of magnetic white dwarf field structures \citep{achilleosetal89-1}.
This model is similar to a centred-dipole, but allows for the origin of the field
to be shifted within the white dwarf. In principle this shift can be applied in three
dimensions, but typically it is only applied along the magnetic field axis by a
fractional amount of the white dwarf radius, $a_z$.
The offset-dipole model has been successfully applied to many different white dwarfs
\citep{achilleosetal92-2,putney+jordan95-1,kulebietal09-1,hollandsetal15-1}
leading to much improved fits with only a single additional free-parameter,
which is advantageous compared to a more general multi-pole expansion.

For a centred-dipole with the magnetic field aligned with the $z$-axis, the
value of the magnetic field at any point on (and external to) the stellar
surface in Cartesian coordinates ($x/y/z$) is given by,
\begin{equation}
    \begin{bmatrix}
    B_x \\ B_y \\ B_z
    \end{bmatrix}
    = \frac{B_d}{2r^5}
    \begin{bmatrix}
    3xz \\ 3yz \\ 3z^2-r^2
    \end{bmatrix}\label{eq:B},
\end{equation}
where $B_{x/y/z}$ are the Cartesian components of the magnetic field, $B_d$ is the polar field strength, 
and $r^2 = x^2 + y^2 + z^2$. The offset-dipole model simply requires making the translation $z\mapsto z-a_z$, in
Equation~\eqref{eq:B} and in the definition of $r^2$. To complete
the offset-dipole model we also allow rotation between the magnetic field axis
and the observer. We implement this by considering coordinate systems for both
the magnetic field and the viewing direction of the observer, with a rotation
matrix used to convert between them.

Using the above model of the white dwarf magnetic field structure, we construct a toy
model spectrum by randomly sampling 10,000 points uniformly across the stellar disc
(i.e. sampled uniformly within the unit circle). For each point on the stellar
disc, $i$, we used Equation~\eqref{eq:B} to calculate the magnetic field vector
(accounting for the chosen inclination).
Then for each transition, $j$, we compute a Zeeman-triplet of three Lorentzian profiles,
using our atomic data from Section~\ref{sec:atoms} to determine their
wavelengths and oscillator strengths. Furthermore, the $\pi$-component is
scaled by a factor $\sin^2\psi/2$, and the $\sigma$-components by a factor
$(1+\cos^2\psi)/4$, which accounts for linear and circular polarisation effects
respectively \citep{putney+jordan95-1}, and where $\psi$ is the angle between
the field line and the observer's line of sight\footnotemark. These three
Lorentzian components are then summed to form an opacity function
\begin{equation}
    \kappa_{ij}(\lambda; B_i, \psi_i) = \sum_{\Delta m_l=-1}^{+1} L_j(\lambda; B_i, \psi_i, \Delta m_l),
\end{equation}
where $L_j$ are the Lorentzian profiles per transition. Finally, the normalised
flux for all transitions at point $i$ is given by
\begin{equation}
    F_i(\lambda; B_i, \psi_i) = \exp\left\{-\sum_{j} A_j \kappa_{ij}(\lambda; B_i, \psi_i)\right\},
\end{equation}
where $A_j$ is a pseudo-abundance which we use to arbitrarily scale the strength of each Zeeman-triplet.
Finally, we compute the integrated flux over the stellar disc as a weighted sum based on the centre-to-limb intensity
of the stellar disc
\begin{equation}
    F(\lambda) = \frac{\sum_i F_i(\lambda; B_i, \psi_i) I(\mu_i)}{\sum_i I(\mu_i)},
\end{equation}
where $I(\mu_i)$ is the intensity across the stellar disc, and where $\mu_i$ is
equivalent to the $z$ coordinate of the $i$-th point on the stellar disc in the
observers frame of reference. We use the logarithmic limb-darkening law for a
6000\,K, $\log g = 8$ white dwarf from \citet{gianninasetal13-1}.

\footnotetext{These oscillator strength scaling factors mean that when the
observer looks down a field line, the $\pi$-component vanishes and the
$\sigma$-components are at maximum intensity, and when the observer looks
perpendicular to a field line the $\pi$-component is at maximum intensity with
the $\sigma$-components at half intensity. In the absence of a
magnetic field where all components overlap, all three scaling factors sum
to one for all angles of $\psi$.}

\subsection{Application to \texorpdfstring{\WD}{SDSSJ1143+6615}}

\begin{figure*}
	\includegraphics[width=\textwidth]{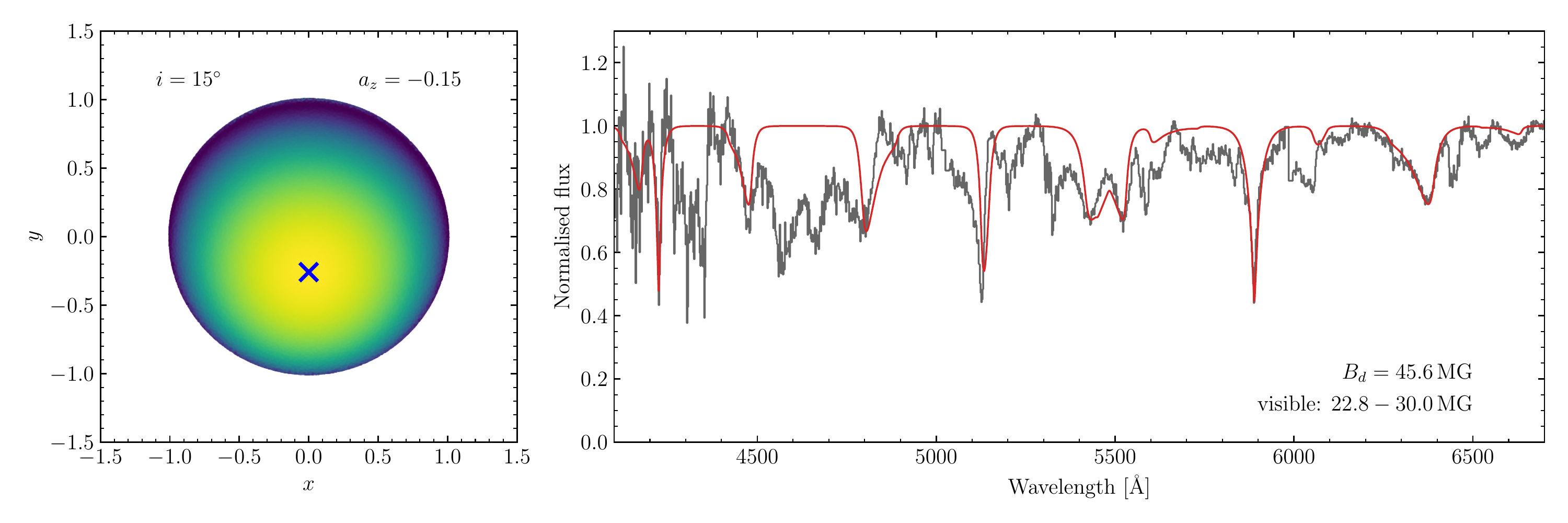}
    \caption{Left: Visualisation of the field structure of \WD\ modeled with an
    offset-dipole. Right: The simulated absorption spectrum of \WD\ (red) using
    data from our finite-field coupled cluster calculations.}
    \label{fig:MWDmodel}
\end{figure*}

We applied the offset dipole model to \WD\ initially focusing on the Na
triplet. From analysing the $\pi$-components of Mg and Na in
Section~\ref{sec:lines}, we established a surface averaged field of $\simeq
30$\,MG, and hence located the features corresponding to the
$\sigma$-components. Due to the asymmetry of the Mg components we decided to
begin our focus on the Na triplet. However, the $\sigma^{-}$ component of Na
and the $\sigma^{+}$ component of Mg are somewhat overlapping ($\simeq
5500$\,\AA), and so we chose to restrict ourselves to the $\pi$ and
$\sigma^{+}$ components of Na ($\simeq 6400$\,\AA). Overall we therefore had
five parameters to adjust: the polar field strength $B_d$, the
dipole-inclination, and the dipole-offset $a_z$, which controlled the field
distribution; plus the Lorentzian line strength ($A_j$ in
Section~\ref{sec:offsetdipole}) and width which are most easily inferred by the
relatively static $\pi$-component.

As described at the start of Section~\ref{sec:magmodel},
the width of the $\sigma^{+}$ component of Na implies a field strength
distribution narrower than the factor of two for a centred dipole.
In the offset-dipole model a narrower distribution can be achieved for negative
values of $a_z$, combined with a low inclination (i.e. viewed close to pole-on).
This implies a wider distribution of field strengths on the opposite hemisphere of the star.
Because $B_d$ in Equation~\eqref{eq:B} no longer corresponds to the field at the poles,
for finite $a_z$, both parameters must be adjusted simultaneously to maintain a polar
field strength of 30\,MG on the visible hemisphere. Manipulating Equation~\ref{eq:B},
and making the substitution $z \mapsto z-a_z$, it can be shown that 
\begin{equation}
    B_d = (1-a_z)^3 B_{z=1},
    \label{eq:Bpole}
\end{equation}
where $B_{z=1}$ is the near-side pole strength of 30\,MG.
Adjusting these parameters by hand\footnotemark, we found good agreement with the shape of
the Na $\sigma^{+}$-component could be achieved with $a_z=-0.15$
(implying $B_d=45.6$\,MG from Equation~\eqref{eq:Bpole})
and a dipole-inclination of $15$\,degrees (Figure~\ref{fig:MWDmodel}).
This also yields a reasonable agreement with the $\sigma^{-}$-component
(at wavelengths where it is not blended with the $\sigma^{+}$-component from Mg).
We then included all other transitions from Section~\ref{sec:atoms}
into the model adjusting only the strengths and widths of the
Lorentzian profiles. A further refinement is required for the \Ion{Mg}{i}
and \Ion{Ca}{i} 6142\,\AA\ triplets as these are $n p\rightarrow (n+1) s$ transitions
(the others are all $n s\rightarrow n p$), and so we scale the component strengths
by Boltzmann factors reflecting the different occupation levels of the lower states.
\footnotetext{While we did attempt a more rigorous least-squares fit to the data, the
lack of a well-defined continuum led to worse results than manual adjustment of the
model parameters.}

Unsurprisingly, the Lorentzian profiles used provide a poor fit for the asymmetric
$\pi$- and $\sigma^{-}$-components of \Ion{Mg}{i}, though reasonable agreement
is found for the $\sigma^{+}$-component. As discussed
previously, this may indicate that the degree of neutral broadening is 
field-dependent, and affects the bluer components more strongly. For the \Ion{Ca}{i}
4227\,\AA\ resonance line, when the width and strength parameters are adjusted
to match the $\pi$-component, the strength and shape of the $\sigma^{+}$-component
($\simeq 4090$\,\AA) also agree well with the observations.
This demonstrates that the
values of $B_d$, $a_z$, and the inclination found from the Na lines are also
appropriate for this transition. For the \Ion{Ca}{ii} triplet, the width of the
$\sigma^{+}$ component is also seen to be in agreement with the data, though
the signal-to-noise ratio in this part of the spectrum is too poor to compare
the shape of the line with the data.
Finally for the \Ion{Ca}{i} 6142\,\AA\ Zeeman-triplet,
only the shape of the $\sigma^{+}$-component in is reasonable agreement with the
data, furthering the argument from Section~\ref{sec:lines} that these transitions
may originate from another source.

\section{Discussion}
\label{sec:discussion}

\subsection{DZs with much stronger fields}

In Section~\ref{sec:magmodel}, we constructed a toy-model for generating simplified
magnetic DZ spectra, including atomic data from Section~\ref{sec:atoms}.
While it turned out that \WD\ has only a 30\,MG field, in principle our model
allows us to generate synthetic spectra for much larger fields,
with 470\,MG covering all the transitions we calculated in Section~\ref{sec:atoms}.
Since ongoing/upcoming spectroscopic surveys such as WEAVE, DESI, SDSS~V, and 4MOST,
are expected to yield hundreds of thousands of white dwarf spectra in the next decade,
we investigate which transitions ought to be focused on for identifying even higher field DZ stars in the future.

\begin{figure*}
	\includegraphics[width=\textwidth]{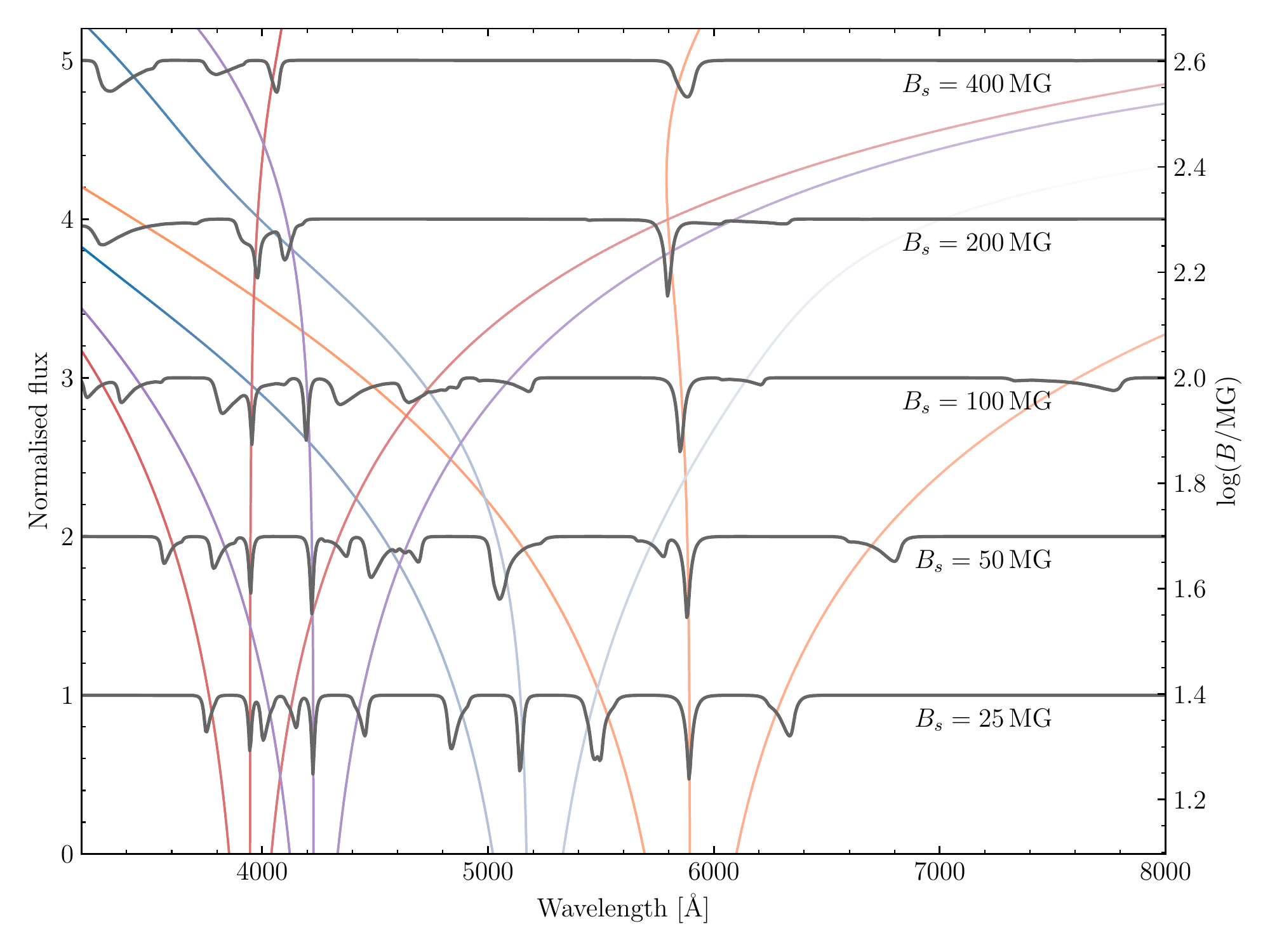}
    \caption{Simulated magnetic DZ spectra for five different surface averaged field
    strengths ($B_s$), with each spectrum offset from one another by 1 in normalised flux.
    The inclination and dipole offset parameters are fixed to the values found for \WD\
    (i.e. 15\,degrees and $-0.15$, respectively).
    The background Zeeman triplets have the same meaning as in Figure\,\ref{fig:spaghetti},
    with the field strength scale given on the right-hand axis.}
    \label{fig:B_range}
\end{figure*}

In Figure~\ref{fig:B_range}, we show models with
average surface fields spanning 25--400\,MG against the same curves
from Figure~\ref{fig:spaghetti}. For all five models, we used the same inclination
and dipole offset as found for \WD, i.e. 15\,degrees and $-0.15$ respectively.
Note that the $B_s$ values are the surface averaged field strengths whereas the
dipole field strength, $B_d$, is approximately 52\,percent larger (see Equation~\ref{eq:Bpole}).

The bottom model has a field $B_s = 25$\,MG, similar in strength to that found
for \WD, and thus shows a similar spectrum. Despite the relatively uniform field
for an inclination of 15\,degrees and $a_z = -0.15$, as the field increases,
the $\sigma$-components still become washed out, and for most of the transitions
are almost invisible at fields of around 100\,MG and above. For \Ion{Mg}{i} the $\sigma^+$
component still remains visible above 100\,MG due to its increase in oscillator
strength.

On the other hand, most of the $\pi$-components remain relatively steady in wavelength.
For the Na $\pi$-component, as already noted in \citet{triplespaper}, the wavelength changes very little below $100$\,MG,
leaving this line similarly sharp as for a 25\,MG field.
The Na line reaches a maximum in blue-shift at 240\,MG
(100\,\AA\ bluer than the field free wavelength), before rapidly turning
around and moving to redder wavelengths. Therefore for $B_s=400$\,MG, the
line becomes much broader, but remains clearly visible. Therefore this
transition ought to be used as a primary marker for identifying cool
magnetic DZ white dwarfs with $>100$\,MG fields.

Similarly the \Ion{Ca}{i} $\pi$-component remains relatively stationary up
to $100$\,MG, but becomes more washed out for larger fields due to the quadratic
Zeeman effect, and becoming broadened to a width of 100\,\AA\ at $400$\,MG.
Therefore this line is likely to be less reliable than the Na $\pi$-component for
identifying the highest field DZs, but will still remain reliable up to 200\,MG.

The \Ion{Ca}{ii} $\pi$-component is also
near stationary, and should still be recognisable even at 400\,MG, making
this a more obvious choice for identifying warmer high field DZs
where the \Ion{Na}{i} and \Ion{Ca}{i} lines may be too weak to identify.
Note that at 300\,MG, the \Ion{Ca}{i} and \Ion{Ca}{ii} $\pi$-components overlap
producing a blended spectral feature.

Finally, the \Ion{Mg}{i} $\pi$-component
experiences a much larger quadratic shift than the other transitions
considered here. Therefore at 400\,MG, the line appears broad and asymmetric
though is notably still visible, in part due to the increased oscillator
strength for this component, which is close to four times larger than
in the field-free case, thereby also showcasing the importance of considering field-dependent intensities. Note that this toy-model does not consider the intrinsic
asymmetry caused by neutral broadening, which itself could be a function
of field strength.

A final consideration is that we have not yet identified all the features in
the spectrum of \WD. Therefore at very high field strengths of 100s of MG,
these unclassified features will also appear shifted into other parts of the spectrum
further complicating the identification of the transitions discussed
above. Furthermore, other strong lines outside the optical such as
the \Ion{Mg}{i} and \Ion{Mg}{ii} resonance lines (field free wavelengths
at 2853\,\AA\ and 2799\,\AA, respectively), may find some of their Zeeman
split components shifted into the optical providing other atomic features
requiring identification.

\subsection{Use in model atmospheres}

Ideally the atomic data we have presented in Section~\ref{sec:atoms} can be
utilised in white dwarf model atmospheres for more detailed analyses of
magnetic DZ stars. As we have shown in this work, however, this is not necessary for a basic assessment.
For simply determining the surface-average field strength, $B_s$,
and which ions are present in the atmosphere, it is sufficient
to simply compare our atomic data with the spectrum in question, as was
demonstrated in Section~\ref{sec:lines} for \WD. Furthermore, determining
the field structure of a white dwarf can be achieved with
a simple model such as the toy-model we demonstrated in Section~\ref{sec:magmodel}.
Importantly our toy-model is computationally efficient,
taking only a few seconds to produce Figure~\ref{fig:MWDmodel}.

Of course, much can still be learned from incorporating
our atomic data into model atmospheres. In our toy-model from Section~\ref{sec:magmodel},
the strength and widths of the Lorentzian profiles we used have no physical basis,
and are simply adjusted to give acceptable agreement with the data.
In a model atmosphere, the strengths and widths of the features
seen in the spectrum of \WD\ can be investigated by adjusting the abundances
and $\Teff$ (and to some extent the surface gravity) of the model, allowing these
atmospheric properties to be measured in a physically meaningful way. 

The main challenge of such an approach is the computation time required. In
the field-free case, the final model spectrum is integrated over the stellar
disc from spectra calculated at different angles between the surface-normal and the observer.
For finite-fields, however, the synthetic spectra must also be calculated over a
grid of field strengths and angles between the field and observer. In particular,
the field strength axis of the grid must be computed at sufficiently fine steps
so that artefacts from undersampling are not present when integrating over the stellar disc.
Therefore, depending on the range of field strengths required, computation
may take hundreds to thousands of times longer than in the field-free case.
If the \Teff, $\log g$, or abundances require refinement when comparing
against a particular spectrum, the grid must then be recomputed with updated
atmospheric parameters, leading to an even larger amount of computation time.

For that reason we have refrained from including our atomic data within model
atmospheres at the present time, and also because it exceeds the scope
of our primary goals of classifying the spectral features of \WD\ and
measuring its field strength. However, future work should perform a detailed
atmospheric analysis of \WD\ utilising the atomic data presented here
to measure its \Teff, $\log g$, and abundances.

\section{Conclusions}
\label{sec:conc}
We have identified \WD\ as DZ white dwarf with strong magnetic field resulting
in its unique spectrum. Using finite-field, coupled-cluster calculations we were
able to identify lines from \Ion{Na}{i}, \Ion{Mg}{i}, and \Ion{Ca}{i--ii} that
were split and shifted by the linear and quadratic Zeeman effects.
This also allowed us to establish a field strength of $\simeq 30$\,MG, demonstrating that
DZ spectra are challenging to interpret at only a few 10\,MG, due to multiple overlapping
transitions from a variety of chemical elements, which is not the case for magnetic DAs or DBs
at the same field strength.
Using the offset-dipole model, we were able to obtain an adequate
fit to the spectral features of Na with an almost pole-on observation angle,
and the dipole offset away from the observer.

Despite our success in elucidating some of the peculiar features in the
spectrum of \WD, many transitions still lack classification at the present
time. Giving consideration to the elements and lines most commonly encountered
in non-magnetic cool DZ stars, future atomic data calculations should
concentrate on Fe and Cr lines, as well as additional transitions of Ca. Because
\WD\ is currently the only available test for these calculations, and only
samples the relatively low-field end, searching for additional high-field DZs
within ongoing and future spectroscopic surveys (such as SDSS\,V, WEAVE, and
DESI) is imperative to test the accuracy of our atomic data further at field
strengths of many 100\,MG.

\section*{Acknowledgements}

MAH was supported by grant ST/V000853/1 from the Science and Technology Facilities Council (STFC).
S.S. acknowledges support by the Deutsche Forschungsgemeinschaft (DFG) grant number STO 1239/1-1 S.S. and within project B5 of the TRR 146 (Project No. 233 630 050).
Based on observations obtained at the international Gemini Observatory, a program of NSF’s NOIRLab, which is managed by the Association of Universities for Research in Astronomy (AURA) under a cooperative agreement with the National Science Foundation on behalf of the Gemini Observatory partnership: the National Science Foundation (United States), National Research Council (Canada), Agencia Nacional de Investigaci\'{o}n y Desarrollo (Chile), Ministerio de Ciencia, Tecnolog\'{i}a e Innovaci\'{o}n (Argentina), Minist\'{e}rio da Ci\^{e}ncia, Tecnologia, Inova\c{c}\~{o}es e Comunica\c{c}\~{o}es (Brazil), and Korea Astronomy and Space Science Institute (Republic of Korea).
For the purpose of open access, the authors has applied a creative commons attribution (CC BY) licence (where permitted by UKRI, ‘open government licence’ or ‘creative commons attribution no-derivatives (CC BY-ND) licence’ may be stated instead) to any author accepted manuscript version arising.

\section*{Data Availability}

All observations of \WD\ are either public (SDSS, Gaia) or no longer
proprietary (Gemini). The Gemini data can be be obtained from the Gemini
Observatory Archive, using program ID GN-2015B-FT-29. Atomic data calculations
presented in this work are available in the appendix tables.



\bibliographystyle{mnras}
\bibliography{aamnem99,aabib,stella} 

\begin{thebibliography}{}
\makeatletter
\relax
\def\mn@urlcharsother{\let\do\@makeother \do\$\do\&\do\#\do\^\do\_\do\%\do\~}
\def\mn@doi{\begingroup\mn@urlcharsother \@ifnextchar [ {\mn@doi@}
  {\mn@doi@[]}}
\def\mn@doi@[#1]#2{\def\@tempa{#1}\ifx\@tempa\@empty \href
  {http://dx.doi.org/#2} {doi:#2}\else \href {http://dx.doi.org/#2} {#1}\fi
  \endgroup}
\def\mn@eprint#1#2{\mn@eprint@#1:#2::\@nil}
\def\mn@eprint@arXiv#1{\href {http://arxiv.org/abs/#1} {{\tt arXiv:#1}}}
\def\mn@eprint@dblp#1{\href {http://dblp.uni-trier.de/rec/bibtex/#1.xml}
  {dblp:#1}}
\def\mn@eprint@#1:#2:#3:#4\@nil{\def\@tempa {#1}\def\@tempb {#2}\def\@tempc
  {#3}\ifx \@tempc \@empty \let \@tempc \@tempb \let \@tempb \@tempa \fi \ifx
  \@tempb \@empty \def\@tempb {arXiv}\fi \@ifundefined
  {mn@eprint@\@tempb}{\@tempb:\@tempc}{\expandafter \expandafter \csname
  mn@eprint@\@tempb\endcsname \expandafter{\@tempc}}}

\bibitem[\protect\citeauthoryear{{Achilleos} \& {Wickramasinghe}}{{Achilleos}
  \& {Wickramasinghe}}{1989}]{achilleosetal89-1}
{Achilleos} N.,  {Wickramasinghe} D.~T.,  1989, \mn@doi [ApJ] {10.1086/168024},
  \href {http://adsabs.harvard.edu/abs/1989ApJ...346..444A} {346, 444}

\bibitem[\protect\citeauthoryear{{Achilleos}, {Wickramasinghe}, {Liebert},
  {Saffer}  \& {Grauer}}{{Achilleos} et~al.}{1992}]{achilleosetal92-2}
{Achilleos} N.,  {Wickramasinghe} D.~T.,  {Liebert} J.,  {Saffer} R.~A.,
  {Grauer} A.~D.,  1992, ApJ, 396, 273

\bibitem[\protect\citeauthoryear{{Allard}, {Leininger}, {Gad{\'e}a},
  {Brousseau-Couture}  \& {Dufour}}{{Allard} et~al.}{2016}]{allardetal16-1}
{Allard} N.~F.,  {Leininger} T.,  {Gad{\'e}a} F.~X.,  {Brousseau-Couture} V.,
  {Dufour} P.,  2016, \mn@doi [A\&A] {10.1051/0004-6361/201527826}, \href
  {http://adsabs.harvard.edu/abs/2016A%26A...588A.142A} {588, A142}

\bibitem[\protect\citeauthoryear{{Angel}, {Liebert}  \& {Stockman}}{{Angel}
  et~al.}{1985}]{angeletal85-1}
{Angel} J. R.~P.,  {Liebert} J.,   {Stockman} H.~S.,  1985, ApJ, \href
  {1985ApJ...292..260A} {292, 260}

\bibitem[\protect\citeauthoryear{{Bagnulo} \& {Landstreet}}{{Bagnulo} \&
  {Landstreet}}{2018}]{bagnulo+landstreet18-1}
{Bagnulo} S.,  {Landstreet} J.~D.,  2018, \mn@doi [A\&A]
  {10.1051/0004-6361/201833235}, \href
  {https://ui.adsabs.harvard.edu/abs/2018A&A...618A.113B} {618, A113}

\bibitem[\protect\citeauthoryear{{Bagnulo} \& {Landstreet}}{{Bagnulo} \&
  {Landstreet}}{2019}]{bagnulo+landstreet19-1}
{Bagnulo} S.,  {Landstreet} J.~D.,  2019, \mn@doi [A\&A]
  {10.1051/0004-6361/201936068}, \href
  {https://ui.adsabs.harvard.edu/abs/2019A&A...630A..65B} {630, A65}

\bibitem[\protect\citeauthoryear{{Bagnulo} \& {Landstreet}}{{Bagnulo} \&
  {Landstreet}}{2021}]{bagnulo+landstreet21-1}
{Bagnulo} S.,  {Landstreet} J.~D.,  2021, \mn@doi [MNRAS]
  {10.1093/mnras/stab2046}, \href
  {https://ui.adsabs.harvard.edu/abs/2021MNRAS.507.5902B} {507, 5902}

\bibitem[\protect\citeauthoryear{{Becken}, {Schmelcher}  \&
  {Diakonos}}{{Becken} et~al.}{1999}]{beckenetal99-1}
{Becken} W.,  {Schmelcher} P.,   {Diakonos} F.~K.,  1999, \mn@doi [Journal of
  Physics B Atomic Molecular Physics] {10.1088/0953-4075/32/6/018}, \href
  {https://ui.adsabs.harvard.edu/abs/1999JPhB...32.1557B} {32, 1557}

\bibitem[\protect\citeauthoryear{{Blouin}}{{Blouin}}{2020}]{blouin20-1}
{Blouin} S.,  2020, \mn@doi [MNRAS] {10.1093/mnras/staa1689}, \href
  {https://ui.adsabs.harvard.edu/abs/2020MNRAS.496.1881B} {496, 1881}

\bibitem[\protect\citeauthoryear{{Blouin}, {Dufour}, {Allard}, {Salim}, {Rich}
  \& {Koopmans}}{{Blouin} et~al.}{2019}]{blouinetal19-1}
{Blouin} S.,  {Dufour} P.,  {Allard} N.~F.,  {Salim} S.,  {Rich} R.~M.,
  {Koopmans} L.~V.~E.,  2019, \mn@doi [ApJ] {10.3847/1538-4357/ab0081}, \href
  {https://ui.adsabs.harvard.edu/abs/2019ApJ...872..188B} {872, 188}

\bibitem[\protect\citeauthoryear{{Brinkworth}, {Burleigh}, {Lawrie}, {Marsh}
  \& {Knigge}}{{Brinkworth} et~al.}{2013}]{brinkworthetal13-1}
{Brinkworth} C.~S.,  {Burleigh} M.~R.,  {Lawrie} K.,  {Marsh} T.~R.,   {Knigge}
  C.,  2013, \mn@doi [ApJ] {10.1088/0004-637X/773/1/47}, \href
  {http://adsabs.harvard.edu/abs/2013ApJ...773...47B} {773, 47}

\bibitem[\protect\citeauthoryear{{Dennihy}, {Debes}, {Dunlap}, {Dufour},
  {Teske}  \& {Clemens}}{{Dennihy} et~al.}{2016}]{dennihyetal16-1}
{Dennihy} E.,  {Debes} J.~H.,  {Dunlap} B.~H.,  {Dufour} P.,  {Teske} J.~K.,
  {Clemens} J.~C.,  2016, \mn@doi [ApJ] {10.3847/0004-637X/831/1/31}, \href
  {http://adsabs.harvard.edu/abs/2016ApJ...831...31D} {831, 31}

\bibitem[\protect\citeauthoryear{{Doyle}, {Young}, {Klein}, {Zuckerman}  \&
  {Schlichting}}{{Doyle} et~al.}{2019}]{doyleetal19-1}
{Doyle} A.~E.,  {Young} E.~D.,  {Klein} B.,  {Zuckerman} B.,   {Schlichting}
  H.~E.,  2019, \mn@doi [Science] {10.1126/science.aax3901}, \href
  {https://ui.adsabs.harvard.edu/abs/2019Sci...366..356D} {366, 356}

\bibitem[\protect\citeauthoryear{{Dufour}, {Bergeron}, {Schmidt}, {Liebert},
  {Harris}, {Knapp}, {Anderson}  \& {Schneider}}{{Dufour}
  et~al.}{2006}]{dufouretal06-1}
{Dufour} P.,  {Bergeron} P.,  {Schmidt} G.~D.,  {Liebert} J.,  {Harris} H.~C.,
  {Knapp} G.~R.,  {Anderson} S.~F.,   {Schneider} D.~P.,  2006, \mn@doi [ApJ]
  {10.1086/508144}, \href {2006ApJ...651.1112D} {651, 1112}

\bibitem[\protect\citeauthoryear{{Dufour}, {Liebert}, {Fontaine}  \&
  {Behara}}{{Dufour} et~al.}{2007}]{dufouretal07-1}
{Dufour} P.,  {Liebert} J.,  {Fontaine} G.,   {Behara} N.,  2007, \mn@doi [Nat]
  {10.1038/nature06318}, \href {2007Natur.450..522D} {450, 522}

\bibitem[\protect\citeauthoryear{{Dufour}, {Fontaine}, {Liebert}, {Schmidt}  \&
  {Behara}}{{Dufour} et~al.}{2008}]{dufouretal08-1}
{Dufour} P.,  {Fontaine} G.,  {Liebert} J.,  {Schmidt} G.~D.,   {Behara} N.,
  2008, \mn@doi [ApJ] {10.1086/589855}, \href {2008ApJ...683..978D} {683, 978}

\bibitem[\protect\citeauthoryear{{Dufour}, {Kilic}, {Fontaine}, {Bergeron},
  {Melis}  \& {Bochanski}}{{Dufour} et~al.}{2012}]{dufouretal12-1}
{Dufour} P.,  {Kilic} M.,  {Fontaine} G.,  {Bergeron} P.,  {Melis} C.,
  {Bochanski} J.,  2012, \mn@doi [ApJ] {10.1088/0004-637X/749/1/6}, \href
  {2012ApJ...749....6D} {749, 6}

\bibitem[\protect\citeauthoryear{{Dufour} et~al.,}{{Dufour}
  et~al.}{2015}]{dufouretal15-1}
{Dufour} P.,  et~al., 2015, in {Dufour} P.,  {Bergeron} P.,   {Fontaine} G.,
  eds,  Astronomical Society of the Pacific Conference Series Vol. 493, 19th
  European Workshop on White Dwarfs. p.~37

\bibitem[\protect\citeauthoryear{{Dunlap} \& {Clemens}}{{Dunlap} \&
  {Clemens}}{2015}]{dunlap+clemens15-1}
{Dunlap} B.~H.,  {Clemens} J.~C.,  2015, in {Dufour} P.,  {Bergeron} P.,
  {Fontaine} G.,  eds,  Astronomical Society of the Pacific Conference Series
  Vol. 493, 19th European Workshop on White Dwarfs. p.~547

\bibitem[\protect\citeauthoryear{{Farihi}, {Dufour}, {Napiwotzki}  \&
  {Koester}}{{Farihi} et~al.}{2011}]{farihietal11-2}
{Farihi} J.,  {Dufour} P.,  {Napiwotzki} R.,   {Koester} D.,  2011, \mn@doi
  [MNRAS] {10.1111/j.1365-2966.2011.18325.x}, \href
  {http://adsabs.harvard.edu/abs/2011MNRAS.413.2559F} {413, 2559}

\bibitem[\protect\citeauthoryear{{Farihi}, {G{\"a}nsicke}  \&
  {Koester}}{{Farihi} et~al.}{2013}]{farihietal13-1}
{Farihi} J.,  {G{\"a}nsicke} B.~T.,   {Koester} D.,  2013, \mn@doi [Science]
  {10.1126/science.1239447}, \href
  {http://adsabs.harvard.edu/abs/2013Sci...342..218F} {342, 218}

\bibitem[\protect\citeauthoryear{{Farihi} et~al.,}{{Farihi}
  et~al.}{2022}]{farihietal22-1}
{Farihi} J.,  et~al., 2022, \mn@doi [MNRAS] {10.1093/mnras/stab3475}, \href
  {https://ui.adsabs.harvard.edu/abs/2022MNRAS.511.1647F} {511, 1647}

\bibitem[\protect\citeauthoryear{{Ferrario}, {Wickramasinghe}  \&
  {Kawka}}{{Ferrario} et~al.}{2020}]{ferrarioetal20-1}
{Ferrario} L.,  {Wickramasinghe} D.,   {Kawka} A.,  2020, \mn@doi [Advances in
  Space Research] {10.1016/j.asr.2019.11.012}, \href
  {https://ui.adsabs.harvard.edu/abs/2020AdSpR..66.1025F} {66, 1025}

\bibitem[\protect\citeauthoryear{{Fontaine}, {Villeneuve}, {Wesemael}  \&
  {Wegner}}{{Fontaine} et~al.}{1984}]{fontaineetal84-1}
{Fontaine} G.,  {Villeneuve} B.,  {Wesemael} F.,   {Wegner} G.,  1984, ApJ
  Lett., \href {1984ApJ...277L..61F} {277, L61}

\bibitem[\protect\citeauthoryear{{Forster}, {Strupat}, {Rosner}, {Wunner},
  {Ruder}  \& {Herold}}{{Forster} et~al.}{1984}]{forsteretal84-1}
{Forster} H.,  {Strupat} W.,  {Rosner} W.,  {Wunner} G.,  {Ruder} H.,
  {Herold} H.,  1984, \mn@doi [Journal of Physics B Atomic Molecular Physics]
  {10.1088/0022-3700/17/7/015}, \href
  {http://adsabs.harvard.edu/abs/1984JPhB...17.1301F} {17, 1301}

\bibitem[\protect\citeauthoryear{{Gaia Collaboration} et~al.,}{{Gaia
  Collaboration} et~al.}{2018}]{gaiaDR2-collab-1}
{Gaia Collaboration} et~al., 2018, \mn@doi [A\&A]
  {10.1051/0004-6361/201833051}, \href
  {https://ui.adsabs.harvard.edu/abs/2018A&A...616A...1G} {616, A1}

\bibitem[\protect\citeauthoryear{{Gaia Collaboration} et~al.,}{{Gaia
  Collaboration} et~al.}{2021}]{gaiaEDR3-collab-1}
{Gaia Collaboration} et~al., 2021, \mn@doi [A\&A]
  {10.1051/0004-6361/202039657}, \href
  {https://ui.adsabs.harvard.edu/abs/2021A&A...649A...1G} {649, A1}

\bibitem[\protect\citeauthoryear{{G{\"a}nsicke}, {Marsh}, {Southworth}  \&
  {Rebassa-Mansergas}}{{G{\"a}nsicke} et~al.}{2006}]{gaensickeetal06-3}
{G{\"a}nsicke} B.~T.,  {Marsh} T.~R.,  {Southworth} J.,   {Rebassa-Mansergas}
  A.,  2006, \mn@doi [Science] {10.1126/science.1135033}, \href
  {2006Sci...314.1908G} {314, 1908}

\bibitem[\protect\citeauthoryear{{G{\"a}nsicke}, {Marsh}  \&
  {Southworth}}{{G{\"a}nsicke} et~al.}{2007}]{gaensickeetal07-1}
{G{\"a}nsicke} B.~T.,  {Marsh} T.~R.,   {Southworth} J.,  2007, \mn@doi [MNRAS]
  {10.1111/j.1745-3933.2007.00343.x}, \href {2007MNRAS.380L..35G} {380, L35}

\bibitem[\protect\citeauthoryear{{G{\"a}nsicke}, {Koester}, {Farihi}, {Girven},
  {Parsons}  \& {Breedt}}{{G{\"a}nsicke} et~al.}{2012}]{gaensickeetal12-1}
{G{\"a}nsicke} B.~T.,  {Koester} D.,  {Farihi} J.,  {Girven} J.,  {Parsons}
  S.~G.,   {Breedt} E.,  2012, \mn@doi [MNRAS]
  {10.1111/j.1365-2966.2012.21201.x}, \href {2012MNRAS.424..333G} {424, 333}

\bibitem[\protect\citeauthoryear{{G{\"a}nsicke}, {Schreiber}, {Toloza},
  {Fusillo}, {Koester}  \& {Manser}}{{G{\"a}nsicke}
  et~al.}{2019}]{gaensickeetal19-1}
{G{\"a}nsicke} B.~T.,  {Schreiber} M.~R.,  {Toloza} O.,  {Fusillo} N. P.~G.,
  {Koester} D.,   {Manser} C.~J.,  2019, \mn@doi [Nat]
  {10.1038/s41586-019-1789-8}, \href
  {https://ui.adsabs.harvard.edu/abs/2019Natur.576...61G} {576, 61}

\bibitem[\protect\citeauthoryear{{Gentile Fusillo} et~al.,}{{Gentile Fusillo}
  et~al.}{2019}]{gentilefusilloetal19-1}
{Gentile Fusillo} N.~P.,  et~al., 2019, \mn@doi [MNRAS]
  {10.1093/mnras/sty3016}, \href
  {https://ui.adsabs.harvard.edu/abs/2019MNRAS.482.4570G} {482, 4570}

\bibitem[\protect\citeauthoryear{{Gentile Fusillo} et~al.,}{{Gentile Fusillo}
  et~al.}{2021}]{gentilefusilloetal21-2}
{Gentile Fusillo} N.~P.,  et~al., 2021, \mn@doi [MNRAS]
  {10.1093/mnras/stab2672}, \href
  {https://ui.adsabs.harvard.edu/abs/2021MNRAS.508.3877G} {508, 3877}

\bibitem[\protect\citeauthoryear{{Gianninas}, {Strickland}, {Kilic}  \&
  {Bergeron}}{{Gianninas} et~al.}{2013}]{gianninasetal13-1}
{Gianninas} A.,  {Strickland} B.~D.,  {Kilic} M.,   {Bergeron} P.,  2013,
  \mn@doi [ApJ] {10.1088/0004-637X/766/1/3}, \href
  {http://adsabs.harvard.edu/abs/2013ApJ...766....3G} {766, 3}

\bibitem[\protect\citeauthoryear{{Green} \& {Liebert}}{{Green} \&
  {Liebert}}{1981}]{green+liebert81-1}
{Green} R.~F.,  {Liebert} J.,  1981, \mn@doi [PASP] {10.1086/130785}, \href
  {https://ui.adsabs.harvard.edu/abs/1981PASP...93..105G} {93, 105}

\bibitem[\protect\citeauthoryear{{Greenstein}, {Henry}  \&
  {Oconnell}}{{Greenstein} et~al.}{1985}]{greensteinetal85-1}
{Greenstein} J.~L.,  {Henry} R.~J.~W.,   {Oconnell} R.~F.,  1985, \mn@doi [ApJ
  Lett.] {10.1086/184427}, \href
  {https://ui.adsabs.harvard.edu/abs/1985ApJ...289L..25G} {289, L25}

\bibitem[\protect\citeauthoryear{{Guidry} et~al.,}{{Guidry}
  et~al.}{2021}]{guidryetal21-1}
{Guidry} J.~A.,  et~al., 2021, \mn@doi [ApJ] {10.3847/1538-4357/abee68}, \href
  {https://ui.adsabs.harvard.edu/abs/2021ApJ...912..125G} {912, 125}

\bibitem[\protect\citeauthoryear{Halkier, Helgaker, Jørgensen, Klopper, Koch,
  Olsen  \& Wilson}{Halkier et~al.}{1998}]{basiscorrection}
Halkier A.,  Helgaker T.,  Jørgensen P.,  Klopper W.,  Koch H.,  Olsen J.,
  Wilson A.~K.,  1998, \mn@doi [Chem. Phys. Lett.]
  {https://doi.org/10.1016/S0009-2614(98)00111-0}, 286, 243

\bibitem[\protect\citeauthoryear{Hampe \& Stopkowicz}{Hampe \&
  Stopkowicz}{2017}]{Hampe2017}
Hampe F.,  Stopkowicz S.,  2017, \mn@doi [J. Chem. Phys.] {10.1063/1.4979624},
  146, 154105

\bibitem[\protect\citeauthoryear{Hampe \& Stopkowicz}{Hampe \&
  Stopkowicz}{2019}]{Hampe2019}
Hampe F.,  Stopkowicz S.,  2019, \mn@doi [J. Chem. Theory Comput.]
  {10.1021/acs.jctc.9b00242}, 15, 4036

\bibitem[\protect\citeauthoryear{Hampe, Stopkowicz, Groß, Kitsaras, Grazioli
  \& Blaschke}{Hampe et~al.}{}]{QCUMBRE}
Hampe F.,  Stopkowicz S.,  Groß N.,  Kitsaras M.-P.,  Grazioli L.,   Blaschke
  S., , {QCUMBRE}, Quantum Chemical Utility enabling Magnetic-field dependent
  investigations Benefitting from Rigorous Electron-correlation treatment, \url
  {qcumbre.org}

\bibitem[\protect\citeauthoryear{Hampe, Gross  \& Stopkowicz}{Hampe
  et~al.}{2020}]{triplespaper}
Hampe F.,  Gross N.,   Stopkowicz S.,  2020, Phys. Chem. Chem. Phys., 22, 23522

\bibitem[\protect\citeauthoryear{{Henry} \& {O'Connell}}{{Henry} \&
  {O'Connell}}{1985}]{henry+oconnell85-1}
{Henry} R. J.~W.,  {O'Connell} R.~F.,  1985, PASP, \href {1985PASP...97..333H}
  {97, 333}

\bibitem[\protect\citeauthoryear{{Hollands}}{{Hollands}}{2017}]{Hollands-thesis}
{Hollands} M.~A.,  2017, PhD thesis, University of Warwick, UK

\bibitem[\protect\citeauthoryear{{Hollands}, {G{\"a}nsicke}  \&
  {Koester}}{{Hollands} et~al.}{2015}]{hollandsetal15-1}
{Hollands} M.~A.,  {G{\"a}nsicke} B.~T.,   {Koester} D.,  2015, \mn@doi [MNRAS]
  {10.1093/mnras/stv570}, \href
  {http://adsabs.harvard.edu/abs/2015MNRAS.450..681H} {450, 681}

\bibitem[\protect\citeauthoryear{{Hollands}, {Koester}, {Alekseev}, {Herbert}
  \& {G{\"a}nsicke}}{{Hollands} et~al.}{2017}]{hollandsetal17-1}
{Hollands} M.~A.,  {Koester} D.,  {Alekseev} V.,  {Herbert} E.~L.,
  {G{\"a}nsicke} B.~T.,  2017, \mn@doi [MNRAS] {10.1093/mnras/stx250}, \href
  {http://adsabs.harvard.edu/abs/2017MNRAS.467.4970H} {467, 4970}

\bibitem[\protect\citeauthoryear{{Hollands}, {G{\"a}nsicke}  \&
  {Koester}}{{Hollands} et~al.}{2018a}]{hollandsetal18-1}
{Hollands} M.~A.,  {G{\"a}nsicke} B.~T.,   {Koester} D.,  2018a, \mn@doi
  [MNRAS] {10.1093/mnras/sty592}, \href
  {http://adsabs.harvard.edu/abs/2018MNRAS.477...93H} {477, 93}

\bibitem[\protect\citeauthoryear{{Hollands}, {Tremblay}, {G{\"a}nsicke},
  {Gentile-Fusillo}  \& {Toonen}}{{Hollands} et~al.}{2018b}]{hollandsetal18-2}
{Hollands} M.~A.,  {Tremblay} P.-E.,  {G{\"a}nsicke} B.~T.,  {Gentile-Fusillo}
  N.~P.,   {Toonen} S.,  2018b, \mn@doi [MNRAS] {10.1093/mnras/sty2057}, \href
  {https://ui.adsabs.harvard.edu/abs/2018MNRAS.480.3942H} {480, 3942}

\bibitem[\protect\citeauthoryear{{Hollands} et~al.,}{{Hollands}
  et~al.}{2020}]{hollandsetal20-1}
{Hollands} M.~A.,  et~al., 2020, \mn@doi [Nat Astron.]
  {10.1038/s41550-020-1028-0}, \href
  {https://ui.adsabs.harvard.edu/abs/2020NatAs...4..663H} {4, 663}

\bibitem[\protect\citeauthoryear{{Hollands}, {Tremblay}, {G{\"a}nsicke},
  {Koester}  \& {Gentile-Fusillo}}{{Hollands} et~al.}{2021}]{hollandsetal21-1}
{Hollands} M.~A.,  {Tremblay} P.-E.,  {G{\"a}nsicke} B.~T.,  {Koester} D.,
  {Gentile-Fusillo} N.~P.,  2021, \mn@doi [Nat Astron.]
  {10.1038/s41550-020-01296-7}, \href
  {https://ui.adsabs.harvard.edu/abs/2021NatAs...5..451H} {5, 451}

\bibitem[\protect\citeauthoryear{{Hollands}, {Tremblay}, {G{\"a}nsicke}  \&
  {Koester}}{{Hollands} et~al.}{2022}]{hollandsetal22-1}
{Hollands} M.~A.,  {Tremblay} P.~E.,  {G{\"a}nsicke} B.~T.,   {Koester} D.,
  2022, \mn@doi [MNRAS] {10.1093/mnras/stab3696}, \href
  {https://ui.adsabs.harvard.edu/abs/2022MNRAS.511...71H} {511, 71}

\bibitem[\protect\citeauthoryear{{Horne}}{{Horne}}{1986}]{horne86-1}
{Horne} K.,  1986, PASP, \href {1986PASP...98..609H} {98, 609}

\bibitem[\protect\citeauthoryear{{Hoskin} et~al.,}{{Hoskin}
  et~al.}{2020}]{hoskinetal20-1}
{Hoskin} M.~J.,  et~al., 2020, \mn@doi [MNRAS] {10.1093/mnras/staa2717}, \href
  {https://ui.adsabs.harvard.edu/abs/2020MNRAS.499..171H} {499, 171}

\bibitem[\protect\citeauthoryear{{Izquierdo}, {Toloza}, {G{\"a}nsicke},
  {Rodr{\'\i}guez-Gil}, {Farihi}, {Koester}, {Guo}  \& {Redfield}}{{Izquierdo}
  et~al.}{2021}]{izquierdoetal21-1}
{Izquierdo} P.,  {Toloza} O.,  {G{\"a}nsicke} B.~T.,  {Rodr{\'\i}guez-Gil} P.,
  {Farihi} J.,  {Koester} D.,  {Guo} J.,   {Redfield} S.,  2021, \mn@doi
  [MNRAS] {10.1093/mnras/staa3987}, \href
  {https://ui.adsabs.harvard.edu/abs/2021MNRAS.501.4276I} {501, 4276}

\bibitem[\protect\citeauthoryear{{Jordan}, {Schmelcher}, {Becken}  \&
  {Schweizer}}{{Jordan} et~al.}{1998}]{jordanetal98-2}
{Jordan} S.,  {Schmelcher} P.,  {Becken} W.,   {Schweizer} W.,  1998, A\&A,
  \href {1998A&A...336L..33J} {336, L33}

\bibitem[\protect\citeauthoryear{{Jura}}{{Jura}}{2003}]{jura03-1}
{Jura} M.,  2003, \mn@doi [ApJ Lett.] {10.1086/374036}, \href
  {2003ApJ...584L..91J} {584, L91}

\bibitem[\protect\citeauthoryear{{Kawka} \& {Vennes}}{{Kawka} \&
  {Vennes}}{2011}]{kawka+vennes11-1}
{Kawka} A.,  {Vennes} S.,  2011, \mn@doi [A\&A] {10.1051/0004-6361/201117078},
  532, A7

\bibitem[\protect\citeauthoryear{{Kawka} \& {Vennes}}{{Kawka} \&
  {Vennes}}{2014}]{kawka+vennes14-1}
{Kawka} A.,  {Vennes} S.,  2014, \mn@doi [MNRAS] {10.1093/mnrasl/slu004}, \href
  {http://adsabs.harvard.edu/abs/2014MNRAS.439L..90K} {439, L90}

\bibitem[\protect\citeauthoryear{{Kawka}, {Vennes}, {Schmidt}, {Wickramasinghe}
   \& {Koch}}{{Kawka} et~al.}{2007}]{kawkaetal07-1}
{Kawka} A.,  {Vennes} S.,  {Schmidt} G.~D.,  {Wickramasinghe} D.~T.,   {Koch}
  R.,  2007, \mn@doi [ApJ] {10.1086/509072}, \href {2007ApJ...654..499K} {654,
  499}

\bibitem[\protect\citeauthoryear{{Kawka}, {Vennes}, {Ferrario}  \&
  {Paunzen}}{{Kawka} et~al.}{2019}]{kawkaetal19-1}
{Kawka} A.,  {Vennes} S.,  {Ferrario} L.,   {Paunzen} E.,  2019, \mn@doi
  [MNRAS] {10.1093/mnras/sty3048}, \href
  {https://ui.adsabs.harvard.edu/abs/2019MNRAS.482.5201K} {482, 5201}

\bibitem[\protect\citeauthoryear{{Kawka}, {Vennes}  \& {Ferrario}}{{Kawka}
  et~al.}{2020}]{kawkaetal20-1}
{Kawka} A.,  {Vennes} S.,   {Ferrario} L.,  2020, \mn@doi [MNRAS]
  {10.1093/mnrasl/slz165}, \href
  {https://ui.adsabs.harvard.edu/abs/2020MNRAS.491L..40K} {491, L40}

\bibitem[\protect\citeauthoryear{{Kemp}, {Swedlund}, {Landstreet}  \&
  {Angel}}{{Kemp} et~al.}{1970}]{kempetal70-1}
{Kemp} J.~C.,  {Swedlund} J.~B.,  {Landstreet} J.~D.,   {Angel} J.~R.~P.,
  1970, \mn@doi [ApJ Lett.] {10.1086/180574}, \href
  {http://adsabs.harvard.edu/abs/1970ApJ...161L..77K} {161, L77}

\bibitem[\protect\citeauthoryear{Kendall, {Dunning, Jr.}  \& Harrison}{Kendall
  et~al.}{1992}]{Dunningaug1}
Kendall R.~A.,  {Dunning, Jr.} T.~H.,   Harrison R.~J.,  1992, J. Chem. Phys.,
  96, 6796

\bibitem[\protect\citeauthoryear{{Kepler} et~al.,}{{Kepler}
  et~al.}{2013}]{kepleretal13-1}
{Kepler} S.~O.,  et~al., 2013, \mn@doi [MNRAS] {10.1093/mnras/sts522}, \href
  {http://adsabs.harvard.edu/abs/2013MNRAS.429.2934K} {429, 2934}

\bibitem[\protect\citeauthoryear{{Kepler} et~al.,}{{Kepler}
  et~al.}{2016}]{kepleretal16-1}
{Kepler} S.~O.,  et~al., 2016, \mn@doi [MNRAS] {10.1093/mnras/stv2526}, \href
  {http://adsabs.harvard.edu/abs/2016MNRAS.455.3413K} {455, 3413}

\bibitem[\protect\citeauthoryear{{Kilic}, {Kosakowski}, {Moss}, {Bergeron}  \&
  {Conly}}{{Kilic} et~al.}{2021}]{kilicetal21-1}
{Kilic} M.,  {Kosakowski} A.,  {Moss} A.~G.,  {Bergeron} P.,   {Conly} A.~A.,
  2021, \mn@doi [ApJ Lett.] {10.3847/2041-8213/ac3b60}, \href
  {https://ui.adsabs.harvard.edu/abs/2021ApJ...923L...6K} {923, L6}

\bibitem[\protect\citeauthoryear{Kitsaras \& Stopkowicz}{Kitsaras \&
  Stopkowicz}{2022a}]{petrosnew}
Kitsaras M.-P.,  Stopkowicz S.,  in prep., 2022a

\bibitem[\protect\citeauthoryear{Kitsaras \& Stopkowicz}{Kitsaras \&
  Stopkowicz}{2022b}]{petrosnew2}
Kitsaras M.-P.,  Stopkowicz S.,  in prep., 2022b

\bibitem[\protect\citeauthoryear{{Klein}, {Jura}, {Koester}, {Zuckerman}  \&
  {Melis}}{{Klein} et~al.}{2010}]{kleinetal10-1}
{Klein} B.,  {Jura} M.,  {Koester} D.,  {Zuckerman} B.,   {Melis} C.,  2010,
  \mn@doi [ApJ] {10.1088/0004-637X/709/2/950}, \href {2010ApJ...709..950K}
  {709, 950}

\bibitem[\protect\citeauthoryear{{Kowalski}}{{Kowalski}}{2010}]{kowalski10-1}
{Kowalski} P.~M.,  2010, \mn@doi [A\&A] {10.1051/0004-6361/201015238}, \href
  {https://ui.adsabs.harvard.edu/abs/2010A&A...519L...8K} {519, L8}

\bibitem[\protect\citeauthoryear{Kramida, {Yu.~Ralchenko}, Reader  \& {and NIST
  ASD Team}}{Kramida et~al.}{2022}]{NIST_ASD}
Kramida A.,  {Yu.~Ralchenko} Reader J.,   {and NIST ASD Team} 2022, {NIST
  Atomic Spectra Database (ver. 5.10), [Online]. Available:
  {\tt{https://physics.nist.gov/asd}} [Tue Oct 25 2022]. National Institute of
  Standards and Technology, Gaithersburg, MD.}

\bibitem[\protect\citeauthoryear{{K{\"u}lebi}, {Jordan}, {Euchner},
  {G{\"a}nsicke}  \& {Hirsch}}{{K{\"u}lebi} et~al.}{2009}]{kulebietal09-1}
{K{\"u}lebi} B.,  {Jordan} S.,  {Euchner} F.,  {G{\"a}nsicke} B.~T.,   {Hirsch}
  H.,  2009, \mn@doi [A\&A] {10.1051/0004-6361/200912570}, \href
  {http://adsabs.harvard.edu/abs/2009A%26A...506.1341K} {506, 1341}

\bibitem[\protect\citeauthoryear{{Landstreet} \& {Bagnulo}}{{Landstreet} \&
  {Bagnulo}}{2019}]{landstreet+bagnulo19-1}
{Landstreet} J.~D.,  {Bagnulo} S.,  2019, \mn@doi [A\&A]
  {10.1051/0004-6361/201936009}, \href
  {https://ui.adsabs.harvard.edu/abs/2019A&A...628A...1L} {628, A1}

\bibitem[\protect\citeauthoryear{{Liebert}}{{Liebert}}{1988}]{liebert88-1}
{Liebert} J.,  1988, \mn@doi [PASP] {10.1086/132322}, \href
  {1988PASP..100.1302L} {100, 1302}

\bibitem[\protect\citeauthoryear{{MacDonald}, {Hernanz}  \& {Jose}}{{MacDonald}
  et~al.}{1998}]{macdonaldetal98-1}
{MacDonald} J.,  {Hernanz} M.,   {Jose} J.,  1998, \mn@doi [MNRAS]
  {10.1046/j.1365-8711.1998.01392.x}, \href {1998MNRAS.296..523M} {296, 523}

\bibitem[\protect\citeauthoryear{{Manser}, {G{\"a}nsicke}, {Gentile Fusillo},
  {Ashley}, {Breedt}, {Hollands}, {Izquierdo}  \& {Pelisoli}}{{Manser}
  et~al.}{2020}]{manseretal20-1}
{Manser} C.~J.,  {G{\"a}nsicke} B.~T.,  {Gentile Fusillo} N.~P.,  {Ashley} R.,
  {Breedt} E.,  {Hollands} M.,  {Izquierdo} P.,   {Pelisoli} I.,  2020, \mn@doi
  [MNRAS] {10.1093/mnras/staa359}, \href
  {https://ui.adsabs.harvard.edu/abs/2020MNRAS.493.2127M} {493, 2127}

\bibitem[\protect\citeauthoryear{{Manser} et~al.,}{{Manser}
  et~al.}{2021}]{manseretal21-1}
{Manser} C.~J.,  et~al., 2021, \mn@doi [MNRAS] {10.1093/mnras/stab2948}, \href
  {https://ui.adsabs.harvard.edu/abs/2021MNRAS.508.5657M} {508, 5657}

\bibitem[\protect\citeauthoryear{{Marsh}}{{Marsh}}{1989}]{marsh89-1}
{Marsh} T.~R.,  1989, PASP, \href {1989PASP..101.1032M} {101, 1032}

\bibitem[\protect\citeauthoryear{Matthews \& Stanton}{Matthews \&
  Stanton}{2016}]{PertTriples}
Matthews D.~A.,  Stanton J.~F.,  2016, \mn@doi [J. Chem. Phys.]
  {10.1063/1.4962910}, 145, 124102

\bibitem[\protect\citeauthoryear{Matthews et~al.,}{Matthews
  et~al.}{2020}]{cfourpaper}
Matthews D.~A.,  et~al., 2020, \mn@doi [J. Chem. Phys.] {10.1063/5.0004837},
  152, 214108

\bibitem[\protect\citeauthoryear{{Paquette}, {Pelletier}, {Fontaine}  \&
  {Michaud}}{{Paquette} et~al.}{1986}]{paquetteetal86-2}
{Paquette} C.,  {Pelletier} C.,  {Fontaine} G.,   {Michaud} G.,  1986, \mn@doi
  [ApJS] {10.1086/191112}, \href
  {http://adsabs.harvard.edu/abs/1986ApJS...61..197P} {61, 197}

\bibitem[\protect\citeauthoryear{{Pelletier}, {Fontaine}, {Wesemael}, {Michaud}
   \& {Wegner}}{{Pelletier} et~al.}{1986}]{pelletieretal86-1}
{Pelletier} C.,  {Fontaine} G.,  {Wesemael} F.,  {Michaud} G.,   {Wegner} G.,
  1986, \mn@doi [ApJ] {10.1086/164410}, \href {1986ApJ...307..242P} {307, 242}

\bibitem[\protect\citeauthoryear{{Putney} \& {Jordan}}{{Putney} \&
  {Jordan}}{1995}]{putney+jordan95-1}
{Putney} A.,  {Jordan} S.,  1995, ApJ, \href {1995ApJ...449..863P} {449, 863}

\bibitem[\protect\citeauthoryear{{Reid}, {Liebert}  \& {Schmidt}}{{Reid}
  et~al.}{2001}]{reidetal01-1}
{Reid} I.~N.,  {Liebert} J.,   {Schmidt} G.~D.,  2001, \mn@doi [ApJ Lett.]
  {10.1086/319481}, \href {http://adsabs.harvard.edu/abs/2001ApJ...550L..61R}
  {550, L61}

\bibitem[\protect\citeauthoryear{{Rocchetto}, {Farihi}, {G{\"a}nsicke}  \&
  {Bergfors}}{{Rocchetto} et~al.}{2015}]{rochettoetal15-1}
{Rocchetto} M.,  {Farihi} J.,  {G{\"a}nsicke} B.~T.,   {Bergfors} C.,  2015,
  \mn@doi [MNRAS] {10.1093/mnras/stv282}, \href
  {http://adsabs.harvard.edu/abs/2015MNRAS.449..574R} {449, 574}

\bibitem[\protect\citeauthoryear{{Roesner}, {Wunner}, {Herold}  \&
  {Ruder}}{{Roesner} et~al.}{1984}]{roesneretal84-1}
{Roesner} W.,  {Wunner} G.,  {Herold} H.,   {Ruder} H.,  1984, \mn@doi [Journal
  of Physics B Atomic Molecular Physics] {10.1088/0022-3700/17/1/010}, \href
  {http://adsabs.harvard.edu/abs/1984JPhB...17...29R} {17, 29}

\bibitem[\protect\citeauthoryear{Schimeczek \& Wunner}{Schimeczek \&
  Wunner}{2014a}]{schimeczek+wunner14-2}
Schimeczek C.,  Wunner G.,  2014a, \mn@doi [Computer Physics Communications]
  {http://dx.doi.org/10.1016/j.cpc.2014.05.005}, 185, 2655

\bibitem[\protect\citeauthoryear{{Schimeczek} \& {Wunner}}{{Schimeczek} \&
  {Wunner}}{2014b}]{schimeczek+wunner14-1}
{Schimeczek} C.,  {Wunner} G.,  2014b, \mn@doi [ApJS]
  {10.1088/0067-0049/212/2/26}, \href
  {http://adsabs.harvard.edu/abs/2014ApJS..212...26S} {212, 26}

\bibitem[\protect\citeauthoryear{{Schmidt}, {West}, {Liebert}, {Green}  \&
  {Stockman}}{{Schmidt} et~al.}{1986}]{schmidtetal86-2}
{Schmidt} G.~D.,  {West} S.~C.,  {Liebert} J.,  {Green} R.~F.,   {Stockman}
  H.~S.,  1986, ApJ, \href {1986ApJ...309..218S} {309, 218}

\bibitem[\protect\citeauthoryear{{Schmidt}, {Latter}  \& {Foltz}}{{Schmidt}
  et~al.}{1990}]{schmidtetal90-1}
{Schmidt} G.~D.,  {Latter} W.~B.,   {Foltz} C.~B.,  1990, ApJ, \href
  {1990ApJ...350..758S} {350, 758}

\bibitem[\protect\citeauthoryear{{Schmidt}, {Allen}, {Smith}  \&
  {Liebert}}{{Schmidt} et~al.}{1996}]{schmidtetal96-2}
{Schmidt} G.~D.,  {Allen} R.~G.,  {Smith} P.~S.,   {Liebert} J.,  1996, ApJ,
  \href {1996ApJ...463..320S} {463, 320}

\bibitem[\protect\citeauthoryear{{Schmidt} et~al.,}{{Schmidt}
  et~al.}{2003}]{schmidtetal03-1}
{Schmidt} G.~D.,  et~al., 2003, ApJ, \href {2003ApJ...595.1101} {595, 1101}

\bibitem[\protect\citeauthoryear{Shavitt \& Bartlett}{Shavitt \&
  Bartlett}{2009}]{ccbook}
Shavitt I.,  Bartlett R.~J.,  2009, Many-Body Methods in Chemistry and Physics.
Cambridge Molecular Science, Cambridge

\bibitem[\protect\citeauthoryear{{Sion}, {Greenstein}, {Landstreet}, {Liebert},
  {Shipman}  \& {Wegner}}{{Sion} et~al.}{1983}]{sionetal83-1}
{Sion} E.~M.,  {Greenstein} J.~L.,  {Landstreet} J.~D.,  {Liebert} J.,
  {Shipman} H.~L.,   {Wegner} G.~A.,  1983, \mn@doi [ApJ] {10.1086/161036},
  \href {http://adsabs.harvard.edu/abs/1983ApJ...269..253S} {269, 253}

\bibitem[\protect\citeauthoryear{Stanton, Gauss, Cheng, Harding, Matthews  \&
  Szalay}{Stanton et~al.}{}]{cfour}
Stanton J.~F.,  Gauss J.,  Cheng L.,  Harding M.~E.,  Matthews D.~A.,   Szalay
  P.~G., , {CFOUR, Coupled-Cluster techniques for Computational Chemistry, a
  quantum-chemical program package.} {W}ith contributions from {A}. {A}sthana,
  {A}.{A}. {A}uer, {R}.{J}. {B}artlett, {U}. {B}enedikt, {C}. {B}erger,
  {D}.{E}. {B}ernholdt, {S}. {B}laschke, {Y}. {J}. {B}omble, {S}. {B}urger,
  {O}. {C}hristiansen, {D}. {D}atta, {F}. {E}ngel, {R}. {F}aber, {J}.
  {G}reiner, {M}. {H}eckert, {O}. {H}eun, {M}. Hilgenberg, {C}. {H}uber,
  {T}.-{C}. {J}agau, {D}. {J}onsson, {J}. {J}us{\'e}lius, {T}. Kirsch,
  {M}.-{P}. {K}itsaras, {K}. {K}lein, {G}.{M}. {K}opper, {W}.{J}. {L}auderdale,
  {F}. {L}ipparini, {J}. {L}iu, {T}. {M}etzroth, {L}.{A}. {M}{\"u}ck, {D}.{P}.
  {O}'{N}eill, {T}. {N}ottoli, {J}. {O}swald, {D}.{R}. {P}rice, {E}.
  {P}rochnow, {C}. {P}uzzarini, {K}. {R}uud, {F}. {S}chiffmann, {W}.
  {S}chwalbach, {C}. {S}immons, {S}. {S}topkowicz, {A}. {T}ajti, {J}.
  {V}{\'a}zquez, {F}. {W}ang, {J}.{D}. {W}atts, {C}. {Z}hang, {X}. {Z}heng, and
  the integral packages {MOLECULE} ({J}. {A}lml{\"o}f and {P}.{R}. {T}aylor),
  {PROPS} ({P}.{R}. {T}aylor), {ABACUS} ({T}. {H}elgaker, {H}.{J}. {A}a.
  {J}ensen, {P}. {J}{\o}rgensen, and {J}. {O}lsen), and {ECP} routines by {A}.
  {V}. {M}itin and {C}. van {W}{\"u}llen. {F}or the current version, see
  http://www.cfour.de.

\bibitem[\protect\citeauthoryear{Stopkowicz}{Stopkowicz}{2017}]{perspectiveCC}
Stopkowicz S.,  2017, \mn@doi [Int. J. Quantum Chem.]
  {https://doi.org/10.1002/qua.25391}, 18, e25391

\bibitem[\protect\citeauthoryear{Stopkowicz, Gauss, Lange, Tellgren  \&
  Helgaker}{Stopkowicz et~al.}{2015}]{Stopkowicz2015}
Stopkowicz S.,  Gauss J.,  Lange K.~K.,  Tellgren E.~I.,   Helgaker T.,  2015,
  \mn@doi [J. Chem. Phys.] {10.1063/1.4928056}, 143, 074110

\bibitem[\protect\citeauthoryear{{Swan}, {Farihi}  \& {Wilson}}{{Swan}
  et~al.}{2019a}]{swanetal19-2}
{Swan} A.,  {Farihi} J.,   {Wilson} T.~G.,  2019a, \mn@doi [MNRAS]
  {10.1093/mnrasl/slz014}, \href
  {https://ui.adsabs.harvard.edu/abs/2019MNRAS.484L.109S} {484, L109}

\bibitem[\protect\citeauthoryear{{Swan}, {Farihi}, {Koester}, {Hollands},
  {Parsons}, {Cauley}, {Redfield}  \& {G{\"a}nsicke}}{{Swan}
  et~al.}{2019b}]{swanetal19-1}
{Swan} A.,  {Farihi} J.,  {Koester} D.,  {Hollands} M.,  {Parsons} S.,
  {Cauley} P.~W.,  {Redfield} S.,   {G{\"a}nsicke} B.~T.,  2019b, \mn@doi
  [MNRAS] {10.1093/mnras/stz2337}, \href
  {https://ui.adsabs.harvard.edu/abs/2019MNRAS.490..202S} {490, 202}

\bibitem[\protect\citeauthoryear{{Vanderbosch} et~al.,}{{Vanderbosch}
  et~al.}{2020}]{vanderboschetal20-1}
{Vanderbosch} Z.,  et~al., 2020, \mn@doi [ApJ] {10.3847/1538-4357/ab9649},
  \href {https://ui.adsabs.harvard.edu/abs/2020ApJ...897..171V} {897, 171}

\bibitem[\protect\citeauthoryear{{Vanderbosch} et~al.,}{{Vanderbosch}
  et~al.}{2021}]{vanderboschetal21-1}
{Vanderbosch} Z.~P.,  et~al., 2021, \mn@doi [ApJ] {10.3847/1538-4357/ac0822},
  \href {https://ui.adsabs.harvard.edu/abs/2021ApJ...917...41V} {917, 41}

\bibitem[\protect\citeauthoryear{{Vanderburg} et~al.,}{{Vanderburg}
  et~al.}{2015}]{vanderburgetal15-1}
{Vanderburg} A.,  et~al., 2015, \mn@doi [Nat] {10.1038/nature15527}, \href
  {http://adsabs.harvard.edu/abs/2015Natur.526..546V} {526, 546}

\bibitem[\protect\citeauthoryear{{Vanderburg} et~al.,}{{Vanderburg}
  et~al.}{2020}]{vanderburgetal20-1}
{Vanderburg} A.,  et~al., 2020, \mn@doi [Nat] {10.1038/s41586-020-2713-y},
  \href {https://ui.adsabs.harvard.edu/abs/2020Natur.585..363V} {585, 363}

\bibitem[\protect\citeauthoryear{{Wickramasinghe} \&
  {Ferrario}}{{Wickramasinghe} \&
  {Ferrario}}{2000}]{wickramasinghe+ferrario00-1}
{Wickramasinghe} D.~T.,  {Ferrario} L.,  2000, PASP, \href
  {2000PASP..112..873W} {112, 873}

\bibitem[\protect\citeauthoryear{{Williams}, {Montgomery}, {Winget}, {Falcon}
  \& {Bierwagen}}{{Williams} et~al.}{2016}]{williamsetal16-1}
{Williams} K.~A.,  {Montgomery} M.~H.,  {Winget} D.~E.,  {Falcon} R.~E.,
  {Bierwagen} M.,  2016, \mn@doi [ApJ] {10.3847/0004-637X/817/1/27}, \href
  {https://ui.adsabs.harvard.edu/abs/2016ApJ...817...27W} {817, 27}

\bibitem[\protect\citeauthoryear{{Wilson}, {G{\"a}nsicke}, {Koester}, {Toloza},
  {Pala}, {Breedt}  \& {Parsons}}{{Wilson} et~al.}{2015}]{wilsonetal15-1}
{Wilson} D.~J.,  {G{\"a}nsicke} B.~T.,  {Koester} D.,  {Toloza} O.,  {Pala}
  A.~F.,  {Breedt} E.,   {Parsons} S.~G.,  2015, \mn@doi [MNRAS]
  {10.1093/mnras/stv1201}, \href
  {http://adsabs.harvard.edu/abs/2015MNRAS.451.3237W} {451, 3237}

\bibitem[\protect\citeauthoryear{Woon \& {Dunning, Jr.}}{Woon \& {Dunning,
  Jr.}}{1995}]{DunningPC}
Woon D.~E.,  {Dunning, Jr.} T.~H.,  1995, J. Chem. Phys., 103, 4572

\bibitem[\protect\citeauthoryear{{Wunner}}{{Wunner}}{1987}]{wunner87-1}
{Wunner} G.,  1987, Mitteilungen der Astronomischen Gesellschaft, \href
  {1987MitAG..70..198W} {70, 198}

\bibitem[\protect\citeauthoryear{{Wyatt}, {Farihi}, {Pringle}  \&
  {Bonsor}}{{Wyatt} et~al.}{2014}]{wyattetal14-1}
{Wyatt} M.~C.,  {Farihi} J.,  {Pringle} J.~E.,   {Bonsor} A.,  2014, \mn@doi
  [MNRAS] {10.1093/mnras/stu183}, \href
  {http://adsabs.harvard.edu/abs/2014MNRAS.439.3371W} {439, 3371}

\bibitem[\protect\citeauthoryear{{Xu}, {Jura}, {Koester}, {Klein}  \&
  {Zuckerman}}{{Xu} et~al.}{2014}]{xuetal14-1}
{Xu} S.,  {Jura} M.,  {Koester} D.,  {Klein} B.,   {Zuckerman} B.,  2014,
  \mn@doi [ApJ] {10.1088/0004-637X/783/2/79}, \href
  {http://adsabs.harvard.edu/abs/2014ApJ...783...79X} {783, 79}

\bibitem[\protect\citeauthoryear{{Zhao}}{{Zhao}}{2018}]{zhao18-1}
{Zhao} L.~B.,  2018, \mn@doi [ApJ] {10.3847/1538-4357/aab4fe}, \href
  {https://ui.adsabs.harvard.edu/abs/2018ApJ...856..157Z} {856, 157}

\bibitem[\protect\citeauthoryear{{Zuckerman} \& {Becklin}}{{Zuckerman} \&
  {Becklin}}{1987}]{zuckerman+becklin87-1}
{Zuckerman} B.,  {Becklin} E.~E.,  1987, \mn@doi [Nat] {10.1038/330138a0},
  \href {1987Natur.330..138Z} {330, 138}

\bibitem[\protect\citeauthoryear{{Zuckerman}, {Koester}, {Melis}, {Hansen}  \&
  {Jura}}{{Zuckerman} et~al.}{2007}]{zuckermanetal07-1}
{Zuckerman} B.,  {Koester} D.,  {Melis} C.,  {Hansen} B.~M.,   {Jura} M.,
  2007, \mn@doi [ApJ] {10.1086/522223}, \href {2007ApJ...671..872Z} {671, 872}

\bibitem[\protect\citeauthoryear{{Zuckerman}, {Koester}, {Dufour}, {Melis},
  {Klein}  \& {Jura}}{{Zuckerman} et~al.}{2011}]{zuckermanetal11-1}
{Zuckerman} B.,  {Koester} D.,  {Dufour} P.,  {Melis} C.,  {Klein} B.,   {Jura}
  M.,  2011, \mn@doi [ApJ] {10.1088/0004-637X/739/2/101}, \href
  {2011ApJ...739..101Z} {739, 101}

\makeatother
\end{thebibliography}




\appendix

\section{Atomic data tables}

\begin{table*}
    \centering
    \caption{\label{tab:data_NaI}
    Atomic data for the \Ion{Na}{i} Zeeman triplet under
    an applied magnetic field. The magnetic field strength $B$, is given in both atomic units and in MG. Wavelengths are given in vacuum form. Oscillator strengths
    are calculated according to Equation~\eqref{oscillator}.
    }
    \begin{tabular}{crrrrccc}
      \hline
      & & \multicolumn{3}{c}{Wavelength [\AA]} & \multicolumn{3}{c}{Oscillator strength} \\
      $B$ [B$_0$] & \multicolumn{1}{c}{$B$ [MG]} & \multicolumn{1}{c}{$\sigma^{-}$} & \multicolumn{1}{c}{$\pi$} & \multicolumn{1}{c}{$\sigma^{+}$} & $\sigma^{-}$ & $\pi$ & $\sigma^{+}$ \\
      \hline
      0.000 & 0.0 & 5894.571 & 5894.571 & 5894.571 & 0.324 & 0.324 & 0.324 \\
      0.004 & 9.4 & 5742.745 & 5894.121 & 6048.521 & 0.332 & 0.324 & 0.316 \\
      0.008 & 18.8 & 5593.316 & 5892.750 & 6204.298 & 0.341 & 0.325 & 0.307 \\
      0.012 & 28.2 & 5446.622 & 5890.503 & 6361.706 & 0.349 & 0.325 & 0.299 \\
      0.016 & 37.6 & 5302.977 & 5887.427 & 6520.591 & 0.358 & 0.325 & 0.291 \\
      0.020 & 47.0 & 5162.697 & 5883.594 & 6680.899 & 0.367 & 0.325 & 0.284 \\
      0.024 & 56.4 & 5026.005 & 5879.071 & 6842.550 & 0.375 & 0.325 & 0.275 \\
      0.028 & 65.8 & 4893.157 & 5873.968 & 7005.660 & 0.383 & 0.325 & 0.268 \\
      0.032 & 75.2 & 4764.303 & 5868.374 & 7170.305 & 0.391 & 0.326 & 0.260 \\
      0.036 & 84.6 & 4639.560 & 5862.401 & 7336.642 & 0.400 & 0.326 & 0.253 \\
      0.040 & 94.0 & 4519.034 & 5856.224 & 7504.982 & 0.408 & 0.326 & 0.246 \\
      0.060 & 141.0 & 3978.205 & 5824.551 & 8382.905 & 0.444 & 0.327 & 0.211 \\
      0.080 & 188.0 & 3532.705 & 5800.229 & 9351.385 & 0.475 & 0.328 & 0.180 \\
      0.100 & 235.1 & 3166.158 & 5790.578 & 10456.342 & 0.501 & 0.328 & 0.152 \\
      0.120 & 282.1 & 2862.247 & 5797.779 & 11753.431 & 0.523 & 0.327 & 0.127 \\
      0.140 & 329.1 & 2607.352 & 5820.542 & 13310.699 & 0.541 & 0.325 & 0.106 \\
      0.160 & 376.1 & 2390.966 & 5855.747 & 15218.789 & 0.557 & 0.324 & 0.087 \\
      0.180 & 423.1 & 2205.210 & 5899.663 & 17609.247 & 0.570 & 0.322 & 0.071 \\
      0.200 & 470.1 & 2044.197 & 5948.392 & 20689.341 & 0.581 & 0.320 & 0.057 \\
      0.220 & 517.1 & 1903.490 & 5998.674 & 24813.862 & 0.590 & 0.319 & 0.045 \\
      0.240 & 564.1 & 1779.628 & 6048.357 & 30631.655 & 0.597 & 0.318 & 0.035 \\
      0.260 & 611.1 & 1669.853 & 6095.463 & 39457.890 & 0.603 & 0.318 & 0.026 \\
      0.280 & 658.1 & 1571.985 & 6139.070 & 54464.836 & 0.608 & 0.319 & 0.018 \\
      0.300 & 705.2 & 1484.230 & 6179.092 & 85612.964 & 0.611 & 0.319 & 0.011 \\
      0.320 & 752.2 & 1405.100 & 6215.124 & 188444.928 & 0.613 & 0.320 & 0.005 \\
      0.340 & 799.2 & 1333.383 & 6247.316 & 1320875.762 & 0.614 & 0.322 & 0.001 \\
      0.360 & 846.2 & 1268.067 & 6275.954 & 153882.824 & 0.615 & 0.324 & 0.005 \\
      0.380 & 893.2 & 1208.306 & 6301.160 & 84066.792 & 0.614 & 0.325 & 0.009 \\
      0.400 & 940.2 & 1153.390 & 6322.883 & 59099.755 & 0.613 & 0.327 & 0.012 \\
      0.420 & 987.2 & 1102.746 & 6341.015 & 46374.747 & 0.611 & 0.329 & 0.015 \\
      0.440 & 1034.2 & 1055.874 & 6355.731 & 38758.544 & 0.609 & 0.331 & 0.017 \\
      0.460 & 1081.2 & 1010.555 & 6366.008 & 35866.670 & 0.608 & 0.333 & 0.017 \\
      0.480 & 1128.2 & 971.855 & 6371.932 & 30295.527 & 0.604 & 0.335 & 0.019 \\
      0.500 & 1175.3 & 934.065 & 6372.374 & 27799.155 & 0.601 & 0.337 & 0.020 \\
      \hline
    \end{tabular}
\end{table*}

\begin{table*}
    \centering
    \caption{\label{tab:data_MgI} Atomic data for the \Ion{Mg}{i} Zeeman triplet under
    an applied magnetic field. Columns have the same meaning as in Table~\ref{tab:data_NaI}.
    }
    \begin{tabular}{crrrrccc}
        \hline
      & & \multicolumn{3}{c}{Wavelength [\AA]} & \multicolumn{3}{c}{Oscillator strength} \\
      $B$ [B$_0$] & \multicolumn{1}{c}{$B$ [MG]} & \multicolumn{1}{c}{$\sigma^{-}$} & \multicolumn{1}{c}{$\pi$} & \multicolumn{1}{c}{$\sigma^{+}$} & $\sigma^{-}$ & $\pi$ & $\sigma^{+}$ \\
        \hline
        0.000 & 0.0 & 5179.597 & 5179.597 & 5179.597 & 0.138 & 0.135 & 0.137 \\
        0.004 & 9.4 & 5061.068 & 5174.347 & 5294.864 & 0.157 & 0.136 & 0.120 \\
        0.008 & 18.8 & 4940.130 & 5158.717 & 5406.135 & 0.177 & 0.137 & 0.105 \\
        0.012 & 28.2 & 4817.641 & 5133.057 & 5512.746 & 0.199 & 0.139 & 0.092 \\
        0.016 & 37.6 & 4694.469 & 5097.935 & 5614.167 & 0.224 & 0.142 & 0.080 \\
        0.020 & 47.0 & 4571.460 & 5054.112 & 5710.030 & 0.250 & 0.146 & 0.070 \\
        0.024 & 56.4 & 4449.419 & 5002.504 & 5800.152 & 0.279 & 0.150 & 0.060 \\
        0.028 & 65.8 & 4329.099 & 4944.155 & 5884.567 & 0.309 & 0.156 & 0.052 \\
        0.032 & 75.2 & 4211.181 & 4880.187 & 5963.528 & 0.341 & 0.163 & 0.045 \\
        0.036 & 84.6 & 4096.271 & 4811.771 & 6037.528 & 0.373 & 0.171 & 0.038 \\
        0.040 & 94.0 & 3984.896 & 4740.081 & 6107.286 & 0.407 & 0.181 & 0.033 \\
        0.060 & 141.0 & 3493.822 & 4370.457 & 6432.470 & 0.556 & 0.249 & 0.013 \\
        0.080 & 188.0 & 3125.114 & 4054.666 & 6862.945 & 0.576 & 0.345 & 0.004 \\
        0.100 & 235.1 & 2863.579 & 3828.322 & 7609.333 & 0.451 & 0.430 & 0.001 \\
        0.120 & 282.1 & 2672.345 & 3666.824 & 8859.285 & 0.307 & 0.479 & 0.000 \\
        0.140 & 329.1 & 2521.311 & 3537.491 & 10899.854 & 0.201 & 0.504 & 0.000 \\
        0.160 & 376.1 & 2394.409 & 3422.271 & 14450.113 & 0.127 & 0.518 & 0.000 \\
        0.180 & 423.1 & 2283.908 & 3313.842 & 21816.811 & 0.073 & 0.529 & 0.000 \\
        0.200 & 470.1 & 2185.904 & 3209.926 & 45703.577 & 0.034 & 0.539 & 0.000 \\
        \hline
    \end{tabular}
\end{table*}

\begin{table*}
    \centering
    \caption{\label{tab:data_CaII} Atomic data for the \Ion{Ca}{ii} Zeeman triplet under
    an applied magnetic field. Columns have the same meaning as in Table~\ref{tab:data_NaI}.
    }
    \begin{tabular}{crrrrccc}
        \hline
      & & \multicolumn{3}{c}{Wavelength [\AA]} & \multicolumn{3}{c}{Oscillator strength} \\
      $B$ [B$_0$] & \multicolumn{1}{c}{$B$ [MG]} & \multicolumn{1}{c}{$\sigma^{-}$} & \multicolumn{1}{c}{$\pi$} & \multicolumn{1}{c}{$\sigma^{+}$} & $\sigma^{-}$ & $\pi$ & $\sigma^{+}$ \\
        \hline
        0.000 & 0.0 & 3946.314 & 3946.314 & 3946.314 & 0.320 & 0.320 & 0.320 \\
        0.004 & 9.4 & 3876.686 & 3946.392 & 4017.343 & 0.326 & 0.320 & 0.314 \\
        0.008 & 18.8 & 3808.443 & 3946.626 & 4089.790 & 0.331 & 0.320 & 0.309 \\
        0.012 & 28.2 & 3741.567 & 3947.016 & 4163.676 & 0.337 & 0.320 & 0.303 \\
        0.016 & 37.6 & 3676.043 & 3947.561 & 4239.019 & 0.342 & 0.320 & 0.297 \\
        0.020 & 47.0 & 3611.854 & 3948.260 & 4315.843 & 0.348 & 0.320 & 0.292 \\
        0.024 & 56.4 & 3548.987 & 3949.111 & 4394.170 & 0.353 & 0.320 & 0.286 \\
        0.028 & 65.8 & 3487.425 & 3950.113 & 4474.025 & 0.358 & 0.321 & 0.280 \\
        0.032 & 75.2 & 3427.153 & 3951.265 & 4555.436 & 0.363 & 0.321 & 0.275 \\
        0.036 & 84.6 & 3368.155 & 3952.564 & 4638.432 & 0.369 & 0.321 & 0.269 \\
        0.040 & 94.0 & 3310.414 & 3954.008 & 4723.043 & 0.374 & 0.321 & 0.263 \\
        0.060 & 141.0 & 3039.968 & 3963.325 & 5171.589 & 0.398 & 0.324 & 0.236 \\
        0.080 & 188.0 & 2798.221 & 3975.903 & 5666.558 & 0.419 & 0.326 & 0.209 \\
        0.100 & 235.1 & 2582.582 & 3991.409 & 6214.523 & 0.436 & 0.331 & 0.184 \\
        0.120 & 282.1 & 2390.297 & 4009.463 & 6823.401 & 0.450 & 0.336 & 0.160 \\
        0.140 & 329.1 & 2218.604 & 4029.513 & 7501.885 & 0.459 & 0.343 & 0.138 \\
        0.160 & 376.1 & 2064.851 & 4050.637 & 8258.054 & 0.462 & 0.351 & 0.118 \\
        0.180 & 423.1 & 1926.562 & 4071.303 & 9096.442 & 0.458 & 0.360 & 0.100 \\
        0.200 & 470.1 & 1801.483 & 4089.169 & 10012.688 & 0.445 & 0.370 & 0.083 \\
        \hline
    \end{tabular}
\end{table*}

\begin{table*}
    \centering
    \caption{\label{tab:data_CaI4227} Atomic data for the \Ion{Ca}{i} 4227\,\AA\ Zeeman triplet under
    an applied magnetic field. Columns have the same meaning as in Table~\ref{tab:data_NaI}.
    }
    \begin{tabular}{crrrrccc}
        \hline
      & & \multicolumn{3}{c}{Wavelength [\AA]} & \multicolumn{3}{c}{Oscillator strength} \\
      $B$ [B$_0$] & \multicolumn{1}{c}{$B$ [MG]} & \multicolumn{1}{c}{$\sigma^{-}$} & \multicolumn{1}{c}{$\pi$} & \multicolumn{1}{c}{$\sigma^{+}$} & $\sigma^{-}$ & $\pi$ & $\sigma^{+}$ \\
        \hline
        0.000 & 0.0 & 4227.920 & 4227.920 & 4227.920 & 0.612 & 0.612 & 0.612 \\
        0.004 & 9.4 & 4148.624 & 4227.625 & 4307.822 & 0.624 & 0.612 & 0.601 \\
        0.008 & 18.8 & 4070.040 & 4226.742 & 4388.241 & 0.635 & 0.612 & 0.589 \\
        0.012 & 28.2 & 3992.294 & 4225.272 & 4469.117 & 0.647 & 0.613 & 0.578 \\
        0.016 & 37.6 & 3915.516 & 4223.223 & 4550.413 & 0.659 & 0.613 & 0.567 \\
        0.020 & 47.0 & 3839.838 & 4220.600 & 4632.124 & 0.670 & 0.613 & 0.557 \\
        0.024 & 56.4 & 3765.388 & 4217.412 & 4714.272 & 0.682 & 0.614 & 0.546 \\
        0.028 & 65.8 & 3692.284 & 4213.667 & 4796.905 & 0.694 & 0.615 & 0.535 \\
        0.032 & 75.2 & 3620.629 & 4209.378 & 4880.090 & 0.705 & 0.615 & 0.524 \\
        0.036 & 84.6 & 3550.512 & 4204.555 & 4963.919 & 0.716 & 0.616 & 0.514 \\
        0.040 & 94.0 & 3482.000 & 4199.209 & 5048.494 & 0.727 & 0.617 & 0.503 \\
        0.060 & 141.0 & 3164.874 & 4165.022 & 5487.059 & 0.778 & 0.621 & 0.450 \\
        0.080 & 188.0 & 2890.269 & 4119.406 & 5966.238 & 0.819 & 0.626 & 0.399 \\
        0.100 & 235.1 & 2654.834 & 4063.819 & 6511.058 & 0.852 & 0.631 & 0.349 \\
        0.120 & 282.1 & 2453.542 & 4000.033 & 7156.578 & 0.876 & 0.637 & 0.301 \\
        0.140 & 329.1 & 2281.503 & 3930.006 & 7957.549 & 0.891 & 0.642 & 0.255 \\
        0.160 & 376.1 & 2134.652 & 3855.464 & 9009.840 & 0.898 & 0.648 & 0.211 \\
        0.180 & 423.1 & 2009.768 & 3777.517 & 10498.296 & 0.895 & 0.655 & 0.168 \\
        0.200 & 470.1 & 1904.166 & 3696.491 & 12819.115 & 0.882 & 0.662 & 0.127 \\
        \hline
    \end{tabular}
\end{table*}

\begin{table*}
    \centering
    \caption{\label{tab:data_CaI6142} Atomic data for the \Ion{Ca}{i} 6142\,\AA\ Zeeman triplet under
    an applied magnetic field. Columns have the same meaning as in Table~\ref{tab:data_NaI}.
    }
    \begin{tabular}{crrrrccc}
        \hline
      & & \multicolumn{3}{c}{Wavelength [\AA]} & \multicolumn{3}{c}{Oscillator strength} \\
      $B$ [B$_0$] & \multicolumn{1}{c}{$B$ [MG]} & \multicolumn{1}{c}{$\sigma^{-}$} & \multicolumn{1}{c}{$\pi$} & \multicolumn{1}{c}{$\sigma^{+}$} & $\sigma^{-}$ & $\pi$ & $\sigma^{+}$ \\
        \hline
        0.000 & 0.0 & 6143.862 & 6143.862 & 6143.862 & 0.149 & 0.149 & 0.149 \\
        0.004 & 9.4 & 5976.357 & 6134.649 & 6306.090 & 0.170 & 0.150 & 0.130 \\
        0.008 & 18.8 & 5805.312 & 6107.587 & 6461.705 & 0.192 & 0.152 & 0.114 \\
        0.012 & 28.2 & 5632.852 & 6063.618 & 6610.116 & 0.216 & 0.156 & 0.098 \\
        0.016 & 37.6 & 5460.710 & 6004.725 & 6750.751 & 0.241 & 0.161 & 0.085 \\
        0.020 & 47.0 & 5290.838 & 5933.170 & 6884.019 & 0.268 & 0.168 & 0.072 \\
        0.024 & 56.4 & 5124.852 & 5851.294 & 7010.699 & 0.295 & 0.175 & 0.062 \\
        0.028 & 65.8 & 4964.118 & 5761.377 & 7132.077 & 0.321 & 0.184 & 0.052 \\
        0.032 & 75.2 & 4809.669 & 5666.004 & 7249.747 & 0.347 & 0.194 & 0.043 \\
        0.036 & 84.6 & 4662.011 & 5567.124 & 7364.999 & 0.372 & 0.205 & 0.036 \\
        0.040 & 94.0 & 4521.526 & 5466.431 & 7479.478 & 0.394 & 0.216 & 0.030 \\
        0.060 & 141.0 & 3928.656 & 4976.764 & 8105.437 & 0.457 & 0.273 & 0.010 \\
        0.080 & 188.0 & 3494.930 & 4562.450 & 8986.627 & 0.436 & 0.325 & 0.003 \\
        0.100 & 235.1 & 3177.291 & 4234.160 & 10395.103 & 0.367 & 0.364 & 0.001 \\
        0.120 & 282.1 & 2935.563 & 3968.106 & 12750.000 & 0.290 & 0.391 & 0.000 \\
        0.140 & 329.1 & 2740.263 & 3737.364 & 16942.725 & 0.223 & 0.410 & 0.000 \\
        0.160 & 376.1 & 2574.017 & 3525.150 & 25704.466 & 0.169 & 0.425 & 0.000 \\
        0.180 & 423.1 & 2428.204 & 3325.838 & 142223.979 & 0.125 & 0.440 & 0.000 \\
        0.200 & 470.1 & 2299.061 & 3139.785 & 498426.017 & 0.087 & 0.457 & 0.000 \\
        \hline
    \end{tabular}
\end{table*}


\bsp	
\label{lastpage}
\end{document}